\documentclass[journal]{IEEEtran} 
\usepackage{cite} 

\usepackage[T1]{fontenc}    
\usepackage[utf8]{inputenc} 
\usepackage{newtxtext,newtxmath} 
\usepackage[english]{babel} 
\usepackage{xcolor}
\definecolor{mylightblue}{RGB}{70,100,180}

\usepackage{amsmath, amsfonts, mathtools} 

\usepackage{amsmath, amsfonts, mathtools}

\usepackage{graphicx}
 
\usepackage{booktabs}  
\usepackage{array}     
\usepackage{tabularx}  
\newcolumntype{Y}{>{\centering\arraybackslash}X} 

\usepackage{float}
\usepackage{lscape}
\usepackage{multirow}
\usepackage{adjustbox}
 
 \usepackage[caption=false,font=footnotesize]{subfig}
\usepackage{placeins}   

\usepackage[ruled,vlined]{algorithm2e}  
\RestyleAlgo{ruled}  

\usepackage[hidelinks]{hyperref} 

\usepackage{textcomp}
\usepackage{verbatim}

\usepackage{stfloats} 

\usepackage{url} 

\addto\captionsenglish{}

\usepackage{balance}
\usepackage[normalem]{ulem}

\title{Uncertainty-Aware Performance Modeling of High-Power Microwave Counter-UAV Engagements}
\author{Muhammad Khalil, \textit{Member, IEEE}, Ke Wang, \textit{Senior Member, IEEE}, and Jinho Choi, \textit{Fellow, IEEE}

\thanks{The authors Khalil and Wang are with the School of Engineering, RMIT University, Melbourne, Australia. Emails:$\,$ muhammad.khalil@rmit.edu.au,  and  ke.wang@rmit.edu.au. The author Choi is with the University of Adelaide, Adelaide, Australia. Email: jinho.choi@adelaide.edu.au}
}
\begin{document}
\maketitle

\begin{abstract}
This paper develops an uncertainty-aware probabilistic framework for assessing the operational performance of high-power microwave (HPM) counter-UAV systems under stochastic target motion, beam-pointing uncertainty, atmospheric propagation, and uncertain target susceptibility. The model couples stochastic UAV kinematics with an effective antenna/boresight jitter formulation and atmospheric propagation to obtain closed-form statistics of the received pulse energy, from which both per-pulse and cumulative model-predicted engagement-effectiveness probabilities are derived. Near-boresight beam shaping is captured by a jitter-to-gain mapping. Slant-range variability arises from integrated acceleration noise, and path loss includes free-space spreading along with gaseous and rain attenuation; target susceptibility is represented by a logistic energy–response model. The received pulse energy is treated as an effective target-level exposure metric rather than the exact energy absorbed by a specific internal electronic component. Closed-form moments and log-normal closures yield an analytically evaluable per-pulse mean effectiveness probability via Gaussian–Hermite quadrature and a cumulative dwell-time expression under the standard independence assumption.

Analytical predictions closely match Monte Carlo simulations generated under the same modeling assumptions across wide parameter ranges. With a representative vulnerable-target threshold of $E_{\mathrm{th}}=10^{-2}\,\mathrm{J}$, the model predicts $\bar{P}_{\mathrm{kill}}\gtrsim 0.4$ per pulse and $P_{\mathrm{kill,tot}}>99\%$ within $\sim 0.1\,\mathrm{s}$ at kHz PRF under the adopted assumptions. For more hardened platforms with $E_{\mathrm{th}}=10^{-1}\,\mathrm{J}$, the model predicts $\bar{P}_{\mathrm{kill}}\approx 2.2\times10^{-4}$ (i.e., $\approx 0.02\%$) and $P_{\mathrm{kill,tot}}\approx 20\%$ after $1\,\mathrm{s}$ at $f_{\mathrm{PRF}}=1\,\mathrm{kHz}$ under the i.i.d.\ pulse assumption. A closed-form sensitivity (elasticity) analysis shows that performance is dominated by slant range ($S_{\bar{R}}\approx -2$), with a strong secondary dependence on aperture diameter and transmit power, while pointing jitter and atmospheric variability are comparatively less influential in the evaluated regimes. Within the stated modeling assumptions, the resulting framework provides useful guidance for HPM counter-UAV system sizing, trade-off analysis, and risk-aware mission planning.
\end{abstract}

\begin{IEEEkeywords}
High-power microwave, Counter-UAV systems, Probabilistic performance assessment, Pointing uncertainty, Atmospheric propagation, Sensitivity analysis
\end{IEEEkeywords}

\section{Introduction}

Unmanned aerial vehicles (UAVs), ranging from small commercial quadcopters to advanced military platforms, have become important tools in modern defense and security operations. Their rapid proliferation has reshaped contemporary warfare by providing state and non-state actors with versatile and cost-effective capabilities for surveillance, targeted strikes, and electronic warfare \cite{Khawaja2022, Wang2021a}. In particular, the use of inexpensive UAVs, often in coordinated swarms, poses a major challenge to traditional defense systems, which are not well suited to engaging many simultaneous threats \cite{Shakhatreh2019}. Recent conflicts have further highlighted this asymmetry, as defenders are often forced to use disproportionately expensive interception measures against low-cost drones.

Despite significant research efforts and defense investments, existing counter-UAV (C-UAV) technologies still face important limitations. Kinetic approaches, such as artillery and missile interceptors, can be effective against isolated targets, but they become costly, ammunition-limited, and operationally inefficient in swarm scenarios, while also increasing the risk of collateral damage \cite{Sherman2025}. Directed-energy systems, such as high-energy lasers, provide precise engagement capabilities, but their effectiveness is strongly limited by atmospheric conditions, environmental interference, and line-of-sight constraints \cite{Javed2024}. In addition, both kinetic and laser-based systems are inherently sequential in operation, which limits their scalability against coordinated swarms \cite{Newsrael2025}. Non-kinetic methods, including radio-frequency jamming and spoofing, are also becoming less reliable as UAVs increasingly employ autonomous navigation and anti-jamming countermeasures \cite{Branco2025}.

In response to these challenges, high-power microwave (HPM) technology has emerged as a promising approach for countering UAV. HPM weapons deliver concentrated bursts of electromagnetic energy capable of coupling directly into UAV's electronic systems, effectively disrupting or destroying their internal components \cite{Benford2024}. Unlike conventional methods that rely on kinetic impact or thermal destruction, HPM weapons neutralize UAV electronics through rapid and non-kinetic mechanisms, significantly reducing the risk of collateral damage \cite{PhysicalSciences2024}. A critical operational advantage of HPM systems is their speed-of-light engagement capability, which allows for near-instantaneous energy delivery to targets once they are within the weapon's operational beam range \cite{Benford2024}. Additionally, the reliance of HPM systems on electrical energy rather than expendable munitions ensures an effectively unlimited operating magazine, subject only to the available power and cooling constraints. This circumstance provides a substantial economic advantage over conventional munitions \cite{Backstrom2004}. Furthermore, the ability of HPM systems to produce broad-area effects by adjusting antenna configurations and pulse parameters can support the simultaneous disruption or degradation of multiple UAV threats, significantly improving their suitability against swarm tactics \cite{Wang2022c}.

Real-world demonstrations by programs such as the United States Air Force Research Laboratory's Tactical High-Power Microwave Operational Responder (THOR) and Raytheon's PHASER have validated the practical feasibility and operational effectiveness of HPM systems against drone swarms \cite{Trevithick2019, Zhang2025}. Despite their promising capabilities, the effective implementation of HPM systems requires overcoming several intricate technical challenges \cite{Hadley2023}. These include uncertainties related to dynamic beam pointing errors, atmospheric attenuation effects due to humidity and precipitation, and significant variability in UAV electronic susceptibilities resulting from differences in shielding, internal layout, and manufacturing quality \cite{Han2019}. Indeed, experimental findings consistently reveal substantial variations in UAV responses to HPM exposure, underscoring the inherent stochastic nature of HPM-UAV interactions \cite{Zhang2020b}. Consequently, deterministic modeling approaches that ignore or overly simplify these uncertainties are inadequate for reliable operational planning and performance assessment.

This paper addresses these challenges by developing an uncertainty-aware probabilistic engagement framework for assessing the operational performance of HPM counter-UAV systems. The target UAV is modeled as a continuous-time stochastic process that captures both nominal flight behavior and evasive maneuver uncertainty. The position--velocity state evolves under random acceleration disturbances, leading to time-varying but statistically tractable slant-range behavior. This enables closed-form moment matching for the received pulse energy.

The proposed framework couples this stochastic kinematic model with three key components: (i) an effective antenna and boresight-jitter model that captures beam-pointing uncertainty, aperture-limited gain, and near-boresight gain degradation; (ii) an atmospheric propagation model that combines free-space spreading with gaseous and rain attenuation under uncertain operating conditions; and (iii) a literature-informed susceptibility-response model that maps the effective received pulse energy to a model-predicted engagement-effectiveness probability. The received pulse energy is therefore treated as an effective target-level exposure metric rather than the exact energy absorbed by a specific internal UAV component. In contrast to deterministic or idealized engagement assessments, the proposed formulation explicitly incorporates operational uncertainty while preserving analytical tractability.

Within this framework, we derive closed-form expressions for the moments of the received pulse energy and evaluate the mean per-pulse engagement-effectiveness probability using Gaussian–Hermite quadrature. We then extend the analysis to cumulative dwell-time performance, including the standard independent-pulse assumption and an effective-sample-size interpretation for correlated pulse trains.

This formulation makes the trade-offs explicit among transmit power, aperture size, pulse width, pulse-repetition frequency, engagement range, atmospheric variability, pointing stability, and target susceptibility. Finally, we perform a normalized sensitivity (elasticity) analysis to identify the parameters exerting the greatest influence on predicted counter-UAV performance. To maintain practical relevance, the model is parameterized using publicly available information on representative HPM platforms, including THOR, CHAMP, and PHASER. To the best of the authors' knowledge, this is among the first unified probabilistic HPM–UAV engagement frameworks to provide closed-form received-energy and engagement-effectiveness statistics, together with Monte Carlo verification under a common set of modeling assumptions.

The remainder of this paper is organized as follows. Section~\ref{sec:Sys-Mod} develops the stochastic HPM counter-UAV engagement model, covering UAV kinematics, pointing-jitter statistics, atmospheric propagation, and the effective energy-to-response mapping. It also derives closed-form expressions for received-energy moments, as well as the per-pulse and cumulative model-predicted engagement-effectiveness probabilities. Section~\ref{sec:Sensitivity-Analysis-of} presents the sensitivity analysis and discusses the resulting system-design trade-offs. Section~\ref{sec:Results } provides numerical verification and case studies parameterized by representative HPM systems. Section~\ref{sec:Conclusion} concludes with key findings, limitations, and implications for HPM counter-UAV system design and risk-aware mission planning.

\section{\label{sec:Sys-Mod}System Model }

The system configuration considered in this study is defined in a Cartesian coordinate system of local geodetic right-hand $\mathcal{F}_{0}=\text{\ensuremath{\left\{ O,x,y,z\right\} }},$ where the origin $O$ corresponds to the location of HPM. The $x$ axis points north, the $y$-axis points east, and the $z$-axis aligns with the
local vertical direction. The phase center of the transmitting aperture is located at the origin, and its initial boresight is aligned with the x-axis. Electronic beam steering dynamically directs the main lobe of the antenna pattern toward the target. The elevation of the HPM station above mean sea level is assumed to be negligible relative to the engagement ranges, allowing its effect to be omitted in both channel modeling and geometric calculations.

\begin{figure}[!t]
    \centering
    \includegraphics[width=3.3in]{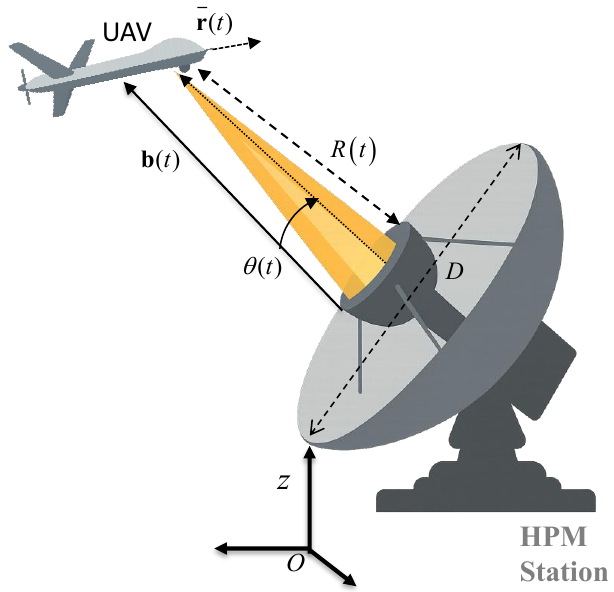}%
    \caption{Engagement geometry between the HPM station and the UAV.}
    \label{fig:1}
\end{figure}

\subsection{\label{subsec:UAV-Kinema}UAV Kinematics and Statistical Modeling
of Range}

The UAV’s movement is represented by a continuous-time stochastic process that captures both nominal flight behavior and rapid evasive maneuvers. At time $t\in$ $[0,T]$, the UAV state is described by a six-dimensional vector $\mathbf{s}(t)=\begin{bmatrix}\mathbf{r}(t)^{\top} & \mathbf{v}(t)^{\top}\end{bmatrix}^{\top}\;\in\;\mathbb{R}^{6}$,
where $\mathbf{r}(t)=\begin{bmatrix}x(t) & y(t) & z(t)\end{bmatrix}^{\!\top}$
represents the position and $\mathbf{v}(t)=\dot{\mathbf{r}}(t)$ denotes the velocity. The time dynamics follow a linear stochastic differential equation,
\begin{equation}
\dot{\mathbf{s}}(t)=\mathcal{\mathbf{F}}\mathbf{s}(t)+\mathbf{w}(t)\in\;\mathbb{R}^{6},\label{eq:S^0}
\end{equation}
where $\mathcal{\mathbf{F}}$ represents the linear kinematic relationship between position and velocity, which is explicitly given by 
\begin{equation}
\mathcal{\mathbf{F}}=\begin{bmatrix}\mathbf{0}_{3} & \mathbf{I}_{3}\\[0.5ex]
\mathbf{0}_{3} & \mathbf{0}_{3}
\end{bmatrix},
\end{equation}
where $\mathbf{0}_{n}$ and $\mathbf{I}_{n}$ denote the $n \times n$ zero matrix and the identity matrix, respectively. This structure ensures that the time derivative of position yields the instantaneous velocity. On the other hand, acceleration is treated as a random process modeled by a white Gaussian process $\mathbf{w}(t)\sim\mathcal{N}\bigl(0,\,\mathbf{Q}\bigr)$  with covariance
\begin{equation}
\mathbb{E}\bigl[\mathbf{w}(t)\,\mathbf{w}^{\top}(\tau)\bigr]=\mathbf{Q}\,\delta(t-\tau),\;\mathbf{Q}=\mathrm{diag}\bigl(\mathbf{0}_{3},\sigma_{a}^{2}\mathbf{I}_{3}\bigr),
\end{equation}
where $\delta(.)$ is the Dirac delta function, and $\sigma_{a}^{2}$ is the one-sided power spectral density of the random acceleration, which represents turbulence and evasive maneuvers. This modeling framework ensures that only the acceleration components of the state vector are affected  by process noise, accurately reflecting the physical sources of uncertainty in UAV dynamics.

By integrating the linear stochastic differential equation \eqref{eq:S^0} with the initial condition \(\mathbf{s}(0)=\bigl[\mathbf{r}_{0}^{\mathsf{T}}\;\mathbf{v}_{0}^{\mathsf{T}}\bigr]^{\mathsf{T}}\in\mathbb{R}^{6}\), we obtain expressions for the mean and covariance of the position vector at time \(t\). The mean position vector is
\begin{equation}
\overline{\mathbf{r}}(t)=\mathbb{E}\bigl[\mathbf{r}(t)\bigr]=\mathbf{r}_{0}+\mathbf{v}_{0}\,t,
\end{equation}
where $\mathbf{r}_{0}$, and $\mathbf{v}_{0}$, are the UAV's initial position and velocity, respectively.

The associated position covariance due to integrated acceleration noise is
\begin{equation}
\mathbf{P}_{rr}(t)=\mathrm{Cov}\bigl[\mathbf{r}(t)\bigr]=\frac{\sigma_{a}^{2}\,t^{3}}{3}\,\mathbf{I}_{3},
\end{equation}
where the factor $\frac{\sigma_{a}^{2}\,t^{3}}{3}$ arises from the integrated effect of white acceleration noise over time. This expression quantifies the growth of uncertainty in each spatial direction due to stochastic accelerations, particularly when the acceleration noise is isotropic. For a general acceleration covariance \(\boldsymbol{\Sigma}_{\mathbf{a}}\), we have \(\mathbf{P}_{rr}(t)=\frac{t^{3}}{3}\boldsymbol{\Sigma}_{\mathbf{a}}\).

Let the instantaneous slant range between the HPM station and the UAV be a random scalar
\begin{equation}
R(t)=\lVert\mathbf{r}(t)\rVert=\sqrt{x^{2}(t)+y^{2}(t)+z^{2}(t)},
\label{eq: R(t)}
\end{equation}

Under the isotropic model $\mathbf{P}_{rr}(t)=\frac{\sigma_{a}^{2}t^{3}}{3}\mathbf{I}_{3}$,
the vector $\mathbf{r}(t)$ is Gaussian with a mean $\overline{\mathbf{r}}(t)$ and covariance
$\sigma_{r}^{2}(t)\mathbf{I}_{3}$, where $\sigma_{r}^{2}(t)=\frac{\sigma_{a}^{2}t^{3}}{3}$.


Irrespective of isotropy, define the squared slant range
\begin{equation}
Q(t)\triangleq R^{2}(t)=\|\mathbf{r}(t)\|^{2}.
\end{equation}
Its first moment is available in closed form:
\begin{equation}
m_{2}(t)\triangleq \mathbb{E}\!\left[Q(t)\right]
=\mathbb{E}\!\left[R^{2}(t)\right]
=\mathrm{tr}\!\bigl(\mathbf{P}_{rr}(t)\bigr)+\|\overline{\mathbf{r}}(t)\|^{2}.
\label{eq: E_R!}
\end{equation}

Moreover, for Gaussian $\mathbf{r}(t)\sim \mathcal{N}(\overline{\mathbf{r}}(t),\mathbf{P}_{rr}(t))$, the variance of $Q(t)=\mathbf{r}^{\top}\mathbf{r}$ is
\[
v_{2}(t)\triangleq \mathrm{Var}\!\left[Q(t)\right]
=2\,\mathrm{tr}\!\bigl(\mathbf{P}_{rr}^{2}(t)\bigr)
+4\,\overline{\mathbf{r}}^{\top}(t)\mathbf{P}_{rr}(t)\overline{\mathbf{r}}(t),
\quad Q(t)=\mathbf r^\top(t)\mathbf r(t).
\]

The exact mean $\mathbb{E}\!\left[R(t)\right]$ does not have an elementary closed form in general. A second-order delta-method expansion of $R(t)=\sqrt{Q(t)}$ around $m_{2}(t)$ yields
\begin{equation}
\bar{R}(t)\triangleq \mathbb{E}\!\left[R(t)\right]
\approx \sqrt{m_{2}(t)}-\frac{v_{2}(t)}{8\,m_{2}^{3/2}(t)}.
\label{eq: R!}
\end{equation}
where $\mathrm{tr}(\cdot)$ denotes the matrix trace and $\|\cdot\|$ denotes the Euclidean norm.

From the same expansion, the range variance is approximated by
\begin{equation}
\sigma_{R}^{2}(t)\approx \frac{v_{2}(t)}{4\,m_{2}(t)}.
\label{eq:Sig}
\end{equation}

For subsequent channel calculations, it is convenient to approximate $R(t)$ as a log-normal random variable when the coefficient of variation is small. Define
\begin{equation}
C_{R}^{2}(t)=\frac{\sigma_{R}^{2}(t)}{\bar{R}^{2}(t)}.
\end{equation}
When $C_{R}^{2}(t)\lesssim 0.2$, we adopt the approximation
$\ln R(t)\sim\mathcal{N}\!\left(\mu_{\ln R}(t),\sigma_{\ln R}^{2}(t)\right)$
with moment-matched parameters
\begin{equation}
\sigma_{\ln R}^{2}(t)=\ln\!\left(1+C_{R}^{2}(t)\right),\:\mu_{\ln R}(t)=\ln\!\bigl(\bar{R}(t)\bigr)-\frac{1}{2}\ln\!\left(1+C_{R}^{2}(t)\right).
\end{equation}

This approximation preserves the first two moments of the slant range and substantially simplifies the ensuing stochastic channel modeling and performance analysis. In particular, it captures both the mean and the spread of the time-varying propagation path length between the transmitter and the UAV, thereby providing a rigorous basis for risk-aware channel characterization and engagement-probability calculations.

\subsection{HPM Transmitter and Beam-Steering Model}

The HPM system employs a circular aperture with a diameter of $D$, an area
of $A=\frac{\pi D^{2}}{4}$, an operating frequency of $f$, a wavelength
of $\lambda=c/f$, and a peak pulse power of $\boldsymbol{P}_{t}$, located
at $O$. 

The boresight direction $\hat{\mathbf{b}}(t)$ is continuously updated
to track the UAV based on real-time measurements and a closed-loop
controller with a bandwidth $f_{trk}$ and latency $\tau_{trk}$.

The instantaneous pointing error relative to the true line-of-sight
is
\begin{equation}
\theta(t)=\cos^{-1}\!\bigl(\hat{\mathbf{b}}(t)\,\cdot\,\hat{\mathbf{r}}(t)\bigr),\quad\hat{\mathbf{r}}(t)=\frac{\mathbf{r}(t)}{\lVert\mathbf{r}(t)\rVert}
\end{equation}

For analytical tractability, the aggregate pointing error is modeled here as an
effective scalar near-boresight angular error,
\[
\theta \sim \mathcal{N}(0,\sigma_\theta^2),
\]
where \(\sigma_\theta^2\) reflects the combined effect of tracking noise,
latency, servo jitter, and related disturbances. This should be interpreted as
a one-dimensional small-angle surrogate for the net boresight misalignment,
rather than as a full two-axis pointing-jitter model. The approximation is most appropriate in the near-boresight regime, where the dominant performance impact is the induced gain reduction.
The far-field power pattern of the aperture near the boresight is approximated by:
\begin{equation}
G(\theta)=G_{0}\exp(-k\,\theta^{2}),
\end{equation}
where
\begin{equation}
G_{0}=\eta\left(\frac{\pi D}{\lambda}\right)^{2},\qquad 0<\eta\le 1,
\end{equation}
is the boresight (maximum) gain, including aperture efficiency $\eta$ (taper/spillover/ohmic losses), and
\begin{equation}
k=\frac{4\ln 2}{\theta_{3\mathrm{dB}}^{2}}
\end{equation}
matches the \emph{full} $3$-dB beamwidth $\theta_{3\mathrm{dB}}$ of the main lobe.
For an ideally uniformly illuminated \emph{circular} aperture, the Airy main lobe gives
\begin{equation}
\theta_{3\mathrm{dB}}\approx 1.03\,\frac{\lambda}{D},
\end{equation}
while in general, $\theta_{3\mathrm{dB}}$ (and $\eta$) can be taken from the measured pattern of the antenna.

Averaging over jitter yields the average effective antenna gain $\bar{G}$, and its variance as follows:
\begin{equation}
\bar{G}=\mathbb{E}\left[G(\theta)\right]=G_{0}\,\bigl(1+2\,k\,\sigma_{\theta}^{2}\bigr)^{-1/2},\label{eq: G!}
\end{equation}
\begin{gather}
\mathrm{Var}\left[G\right]=G_{0}^{2}\Bigl[(1+4\,k\,\sigma_{\theta}^{2})^{-1/2}-(1+2\,k\,\sigma_{\theta}^{2})^{-1}\Bigr].
\end{gather}
This result reveals a critical trade-off between aperture size $D$ and pointing accuracy $\sigma_{\theta}$ . It highlights
that larger antenna sizes, while providing higher potential gain,
become ineffective if the pointing jitter variance is not simultaneously
reduced. Practically, there are diminishing returns when the aperture size exceeds a critical value \textbf{$D_{\mathrm{crit}}\approx \frac{0.437\,\lambda}{\sigma_{\theta}}$.}

\noindent\textit{Accuracy of the Gaussian mainlobe approximation:}
The Gaussian form in \eqref{eq: G!} provides a closed-form mapping from pointing jitter to the mean gain.
However, for a uniformly illuminated circular aperture, the normalized power pattern is the Airy form
\begin{equation}
\frac{G_{\mathrm{Airy}}(\theta)}{G_0}=
\left[\frac{2J_1(u)}{u}\right]^2,\qquad
u=\frac{\pi D}{\lambda}\sin\theta\approx \frac{\pi D}{\lambda}\theta,
\end{equation}
with $J_1(\cdot)$ the first-order Bessel function.
To bound the modeling error as a function of jitter, we compare the Gaussian-based mean gain
$\mathbb{E}[G(\theta)]$ against the Airy-based mean gain $\mathbb{E}[G_{\mathrm{Airy}}(\theta)]$ under $\theta\sim\mathcal{N}(0,\sigma_\theta^2)$, and we report the relative mismatch
\begin{equation}
\epsilon_{G}(\sigma_{\theta})\triangleq\frac{\left|\mathbb{E}\!\left[G(\theta)\right]-\mathbb{E}\!\left[G_{\mathrm{Airy}}(\theta)\right]\right|}{\mathbb{E}\!\left[G_{\mathrm{Airy}}(\theta)\right]},\qquad\theta\sim\mathcal{N}(0,\sigma_{\theta}^{2}).\label{eq:epsilonG_def}
\end{equation}
This comparison, or measured-pattern substitution when available, quantifies the validity range of the
Gaussian approximation versus $\sigma_\theta$ and is reported in the verification section.

For the baseline case (\(\sigma_\theta=1\) mrad and \(\theta_{3\mathrm{dB}}\approx 84\) mrad), we have \(\sigma_\theta/\theta_{3\mathrm{dB}}\ll 0.5\); therefore, the near-boresight Gaussian approximation is expected to be accurate; the quantitative mismatch is reported in Sec.~IV.

In addition, the present jitter model should be understood as an effective scalar approximation of near-boresight misalignment. A more complete treatment could model azimuth and elevation-axis errors jointly; such an extension is
left for future work.

\subsection{Atmospheric Channel and Attenuation Modeling}

The propagation path from the transmitter to the UAV traverses the atmospheric
channel, characterized by two distinct media: (i) free space, in which spherical
spreading dominates, and (ii) the lower troposphere, where molecular absorption
and hydrometeors impose additional attenuation. The free-space path loss obeys
the Friis relation
\begin{equation}
\mathcal{L}_{\mathrm{FS}}(t)=\left(\frac{4\pi R(t)}{\lambda}\right)^{2},
\end{equation}
where \(R(t)\) is the stochastic slant range derived in
Section~\ref{subsec:UAV-Kinema} and \(\lambda=c/f\) is the carrier wavelength.
Since \(\ln R(t)\) is approximately Gaussian, \(\ln \mathcal{L}_{\mathrm{FS}}(t)\)
is also approximately Gaussian, implying that \(\mathcal{L}_{\mathrm{FS}}(t)\)
is approximately log-normal. Closed-form first and second moments therefore
follow directly from the range statistics in \eqref{eq: R!}--\eqref{eq:Sig}
\cite{Goldsmith2005}.

Atmospheric specific attenuation (dB/km) includes gaseous absorption $\gamma_{\mathrm{gas}}$ and precipitation attenuation $\gamma_{\mathrm{rain}}$.
In the present work, the rain specific attenuation is modeled using the ITU-R form
$\gamma_{\mathrm{rain}}=k(f)R_{\mathrm{rain}}^{\alpha(f)}$, while the rain-layer height and rain-rate statistics are interpreted consistently within the broader ITU slant-path framework. To retain analytical tractability, we use a finite-layer slant-path truncation instead of a full site-specific long-term rain-attenuation implementation. Accordingly, the total atmospheric attenuation (in dB) is modeled as
\begin{equation}
A_{\mathrm{atm}}(f,t)=\underbrace{\gamma_{\mathrm{gas}}(f,\mathtt{p},\mathtt{T},\mathit{\mathtt{e}})\,\frac{R(t)}{1000}}_{\textrm{molecular }(\mathrm{O}_{2}+\mathrm{H}_{2}\mathrm{O})}\;+\underbrace{\gamma_{\mathrm{rain}}\!\left(f,R_{\mathrm{rain}}(t)\right)\,\frac{L_{\mathrm{rain}}(t)}{1000}}_{\underset{(\textrm{specificattenuationfromITU-RP.838})}{\textrm{precipitation }}},\label{eq: Att}
\end{equation}
where $R(t)$ is the slant range (m), $\varphi(t)$ is the elevation angle, $\mathtt{p}$ is the total pressure, $\mathtt{T}$ is the absolute temperature, and $\mathit{\mathtt{e}}$ is the water-vapor partial pressure. 

\noindent\textit{Meteorological inputs and baseline conditions:}
Unless otherwise stated, the gaseous specific attenuation $\gamma_{\mathrm{gas}}(f,p,T,e)$
is evaluated using the ITU-R model with fixed near-ground atmospheric conditions
(i.e., $p$, $T$, and $e$ held constant over the short engagement interval), so that
$\gamma_{\mathrm{gas}}$ is quasi-deterministic during the engagement. For precipitation, the instantaneous rain rate $R_{\mathrm{rain}}(t)$ is modeled statistically and enters
$\gamma_{\mathrm{rain}}(f,R_{\mathrm{rain}}(t))$ through the ITU-R specific-attenuation form
$\gamma_{\mathrm{rain}}=k(f)R_{\mathrm{rain}}^{\alpha(f)}$.
The rain-layer height $h_r$ represents the effective precipitation layer, and the in-rain path length is modeled by
$L_{\mathrm{rain}}(t)=\min\{R(t),\,h_r/\sin\varphi(t)\}$.
This combination should be interpreted as a tractable finite-layer approximation to the slant-path geometry rather than a full site-specific ITU long-term attenuation prediction procedure.

For the numerical results in Sec.~IV, rather than committing to a specific site/season dataset, we
parameterize the integrated atmospheric attenuation $A_{\mathrm{atm}}(f,t)$ (dB) by its moments
$(\mu_A,\sigma_A^2)$ (equivalently, by a representative mean specific attenuation
$\bar{\gamma}_{\mathrm{atm}}$ in dB/km, along with the nominal range). Specifically, for a mean range
$\bar{R}$ (m) we use $\mu_A=\bar{\gamma}_{\mathrm{atm}}\,\bar{R}/1000$. Unless otherwise stated, the
baseline uses $f=2.45$~GHz and $\bar{\gamma}_{\mathrm{atm}}=0.2$~dB/km (Table~II), which corresponds to
$\mu_A=0.2$~dB at $\bar{R}=1$~km; we then sweep $(\mu_A,\sigma_A^2)$ to emulate different frequencies and weather severities.


To model precipitation loss with a finite rain layer, we define the effective line-of-sight path length that lies inside the rain medium as $L_{\mathrm{rain}}(t)$ (m). 
Assuming a rain-layer height $h_r$ (m), this in-rain path length is
\begin{equation}
L_{\mathrm{rain}}(t)=\min\!\left\{R(t),\;\frac{h_{r}}{\sin\varphi(t)}\right\},
\label{eq:Lrain}
\end{equation}
which replaces the horizontal-projection approximation (e.g., $R(t)\cos\varphi(t)$).

This expression should be interpreted as a finite-layer slant-path truncation that captures the geometric portion of the line of sight lying inside the rain layer in a manner consistent with the slant-path reasoning used in the ITU-R framework for an airborne endpoint below the rain layer. Optional path-reduction factors to represent horizontally non-uniform rain cells may be incorporated through
$L_{\mathrm{rain,eff}}=r_f\,L_{\mathrm{rain}}$, with $0<r_f\le 1$, when local climatological parameters are available. The present formulation is not intended to represent a full site-specific implementation of the complete long-term ITU slant-path rain-attenuation prediction sequence.


\noindent\textbf{Rain-layer coverage assumption:}
Equation~\eqref{eq:Lrain} models precipitation as a horizontally uniform rain layer of height $h_r$ along the line of sight. Thus, only the segment of the air-to-ground path that lies inside this layer contributes to rain attenuation. At higher elevation angles, the in-rain distance is limited by the layer depth ($\approx h_r/\sin\varphi$), whereas for short-range links, the entire path may lie below $h_r$ and $L_{\mathrm{rain}}(t)=R(t)$. More detailed ITU-R slant-path procedures may include additional reduction factors to capture non-uniform rain cells; neglecting such factors here prioritizes analytical tractability and typically yields a conservative (slightly higher) rain loss for a given $R_{\mathrm{rain}}$.

The current rain-layer approximation is most reliable for LoS links with moderate elevation angles and horizontally homogeneous rain over the engagement footprint. The present formulation is therefore best viewed as a tractable geometric truncation of the rain-affected path, suitable for deriving closed-form attenuation moments, while a full site-specific ITU implementation would additionally require explicit long-term climatological inputs and slant-path reduction steps.


\noindent\textbf{Statistical assumptions and coupling:}
Over the short engagement interval, we treat the near-ground atmospheric state $(p,T,e)$ and the
rain-rate statistics as stationary. Accordingly, $\gamma_{\mathrm{gas}}(f,p,T,e)$ is modeled as quasi-deterministic
for fixed $(p,T,e)$, while $\gamma_{\mathrm{rain}}(f,R_{\mathrm{rain}}(t))$ varies with the random rain-rate process
$R_{\mathrm{rain}}(t)$ and is assumed to be independent of the UAV kinematic state for analytical tractability.
We further neglect any residual dependence between the meteorological drivers of
$\gamma_{\mathrm{gas}}$ and $\gamma_{\mathrm{rain}}$ over the engagement window.

Importantly, even under these independence assumptions, the \emph{total} attenuation
$A_{\mathrm{atm}}(f,t)$ still varies jointly with the UAV geometry through the slant range $R(t)$ and the
elevation-dependent in-rain path length $L_{\mathrm{rain}}(t)$ in \eqref{eq: Att}--\eqref{eq:Lrain}.
If weather–geometry or gas–rain correlations are non-negligible, additional covariance terms would enter
$\sigma_A^2$. This would primarily broaden the received-energy distribution through the second-order moments 
rather than substantially changing its nominal mean and could shift the sensitivity ranking toward more
propagation-dominated behavior under severe-weather or higher-frequency conditions.
Accordingly, the baseline sensitivity conclusions in Sec.~IV-H should be interpreted concerning the
adopted operating point $(\mu_A,\sigma_A^2)$ and may change when atmospheric attenuation becomes more severe.

To capture day-to-day and intra-storm variability, the instantaneous rain rate is modeled as a Gamma random variable:
\begin{equation}
R_{\mathrm{rain}}(t)\sim \mathrm{Gamma}(\alpha_r,\beta_r),
\label{eq: Rrin}
\end{equation}
where \(\alpha_r\) and \(\beta_r\) are the shape and rate parameters obtained from long-term local meteorological records \cite{RadiocommunicationSector2017,IRSR2022}.

Under this assumption, let
\[
\gamma_{\mathrm{rain}}=k(f)\,R_{\mathrm{rain}}^{\alpha(f)},
\]
where \(k(f)\) and \(\alpha(f)\) are the frequency-dependent ITU-R coefficients. Since
\[
\mathbb{E}\!\left[R_{\mathrm{rain}}^{\,s}\right]
=
\frac{\Gamma(\alpha_r+s)}{\beta_r^{\,s}\Gamma(\alpha_r)},
\qquad s>-\alpha_r,
\]
the mean and variance of \(\gamma_{\mathrm{rain}}\) are
\begin{gather}
\mathbb{E}\!\left[\gamma_{\mathrm{rain}}\right]
=
k(f)\,
\frac{\Gamma\!\left(\alpha_r+\alpha(f)\right)}
{\beta_r^{\alpha(f)}\,\Gamma(\alpha_r)},
\nonumber\\
\mathrm{Var}\!\left[\gamma_{\mathrm{rain}}\right]
=
k(f)^2
\left[
\frac{\Gamma\!\left(\alpha_r+2\alpha(f)\right)}
{\beta_r^{2\alpha(f)}\,\Gamma(\alpha_r)}
-
\left(
\frac{\Gamma\!\left(\alpha_r+\alpha(f)\right)}
{\beta_r^{\alpha(f)}\,\Gamma(\alpha_r)}
\right)^2
\right].
\label{eq: y(rain)}
\end{gather}

Because $\gamma_{\mathrm{gas}}$ is quasi-deterministic on short time-scales, and assuming
$\gamma_{\mathrm{rain}}$ is independent of the kinematic range statistics, we approximate
$A_{\mathrm{atm}}(f,t)\sim\mathcal{N}(\mu_{A},\sigma_{A}^{2})$ in dB with
\begin{equation}
\mu_{A}=
\gamma_{\mathrm{gas}}\,\frac{\bar{R}}{1000}
+\mathbb{E}\!\left[\gamma_{\mathrm{rain}}\right]\frac{\bar{L}_{\mathrm{rain}}}{1000},
\label{eq: Mu A}
\end{equation}
\begin{equation}
\sigma_{A}^{2}\approx
\gamma_{\mathrm{gas}}^{2}\left(\frac{\sigma_{R}}{1000}\right)^{2}
+\left(\frac{\bar{L}_{\mathrm{rain}}}{1000}\right)^{2}\mathrm{Var}\!\left(\gamma_{\mathrm{rain}}\right)
+\left(\mathbb{E}\!\left[\gamma_{\mathrm{rain}}\right]\right)^{2}\left(\frac{\sigma_{L}}{1000}\right)^{2},
\label{eq: Sig-A}
\end{equation}
where $\bar{R}=\mathbb{E}[R(t)]$ and $\sigma_{R}^{2}=\mathrm{Var}(R(t))$ denote the mean and variance of the slant range (m), and $\bar{L}_{\mathrm{rain}}=\mathbb{E}[L_{\mathrm{rain}}(t)]$ and $\sigma_{L}^{2}=\mathrm{Var}(L_{\mathrm{rain}}(t))$ denote the mean and variance of the rain-path length (m). Here, $\gamma_{\mathrm{gas}}$ and $\gamma_{\mathrm{rain}}$ are specific attenuations in dB/km. The factor $1000$ converts lengths from meters to kilometers to match the ITU-R attenuation units. The elevation angle is $\bar{\varphi}=\sin^{-1}\!\left(\bar{z}/\bar{R}\right)$, and under a finite rain-layer model with a rain height of $h_r$, one may approximate $\bar{L}_{\mathrm{rain}} \approx \min\{\bar{R},\ h_r/\sin\bar{\varphi}\}$ (and set $\sigma_L^2\approx 0$ for analytical tractability, when needed) \cite{IRSR2022,IRSR2005,Papoulis2002}.

If $\gamma_{\mathrm{gas}}$ and $\gamma_{\mathrm{rain}}$ are correlated, include $2\,\mathrm{Cov}(\gamma_{\mathrm{gas}}R/1000,\gamma_{\mathrm{rain}}L_{\mathrm{rain}}/1000)$ in $\sigma_A^2$.

Equations \eqref{eq: Att}--\eqref{eq: Sig-A} yield tractable closed-form approximations for the first two moments of atmospheric attenuation, which can be incorporated directly into the subsequent energy and engagement-effectiveness probability calculations. In the present manuscript, these moments are derived from the finite-layer slant-path approximation in \eqref{eq:Lrain}, combined with the ITU-R specific-attenuation law for rain. A fuller site-specific ITU treatment could instead first generate attenuation statistics from the complete slant-path prediction sequence and then map those statistics to the effective moments $(\mu_A,\sigma_A^2)$ used here. The adopted formulation avoids computationally intensive weather realizations while retaining a physically consistent geometric treatment of the rain-affected path.

\subsubsection{Large-scale fading and blockage (scope)}
This work focuses on line-of-sight (LoS) dominated HPM–UAV engagements and models propagation losses
through FSPL and ITU-R-based molecular/rain attenuation. In low-altitude air-to-ground channels, additional
large-scale shadow fading due to buildings, terrain, and airframe-related obstructions can be significant and may
reduce the generality of purely LoS-based predictions. Recent UAV air-to-ground measurement studies report
non-negligible shadowing and environment/airframe-dependent loss statistics (e.g., built-up areas and airframe effects in UWB bands \cite{Ni_TITS_2024_U2G_Shadowing}, and fixed-wing rural measurements at 2.7~GHz \cite{Lyu_TAP_2024_27GHz_Rural}).

To keep the closed-form framework intact, shadow fading can be incorporated as an additive dB-domain random
variable.
\begin{equation}
X_{\mathrm{SF}}\sim\mathcal{N}(0,\sigma_{\mathrm{SF}}^2),\qquad
A_{\mathrm{tot}}(f,t)=A_{\mathrm{atm}}(f,t)+X_{\mathrm{SF}},
\end{equation}
which yields a log-normal multiplicative factor in the received energy. In the log-energy domain, this simply
augments the variance:
\begin{equation}
\sigma_{A,\mathrm{tot}}^2=\sigma_A^2+\sigma_{\mathrm{SF}}^2,
\end{equation}
(and equivalently adds $(\ln 10/10)^2\sigma_{\mathrm{SF}}^2$ to the variance of $\ln E_r$, while leaving the
mean loss unchanged.) Therefore, all moment-based expressions in Sec.~III extend directly by replacing
$(\mu_A,\sigma_A^2)$ with $(\mu_A,\sigma_{A,\mathrm{tot}}^2)$.

Blockage/NLoS conditions can be captured (when needed) by a two-state mixture model with an additional excess loss in NLoS and a LoS probability that depends on the elevation angle and the environment. Such extensions
are standard in cellular UAV channel models (e.g., 3GPP aerial UE studies); however, they are outside the primary scope
of the present LoS-focused analysis.

The practical implication of this extension is that large-scale shadow fading primarily broadens the distribution of the received log-energy and, therefore, increases uncertainty in the predicted engagement-effectiveness probability while leaving the main analytical structure unchanged. Accordingly, the baseline results reported in this paper should be interpreted as the most representative of LoS-dominated engagement scenarios. In more heavily cluttered or persistently blocked environments, stronger shadowing and blockage generally reduce the robustness and generality of the predicted engagement effectiveness, and a richer site-specific propagation model may be required.

 \subsection{Received Energy and Probabilistic Engagement-Effectiveness Metric} 

During an engagement, the HPM transmitter illuminates the target for
a dwell time of $T>0$ {[}s{]}. It emits identical rectangular pulses
of peak envelope power$P_{t}$ {[}W{]} and duration $\tau_{p}$ {[}s{]}
at a pulse-repetition frequency $f_{PRF}${[}$\mathrm{s^{-1}}${]}. The number of pulses within the dwell is
\begin{equation}
N=f_{PRF}\,T.
\end{equation}
and, by design, $\tau_{p}\ll1/f_{PRF}$, the duty cycle $\delta=\tau_{p}\,f_{PRF}$
remains small, ensuring thermal feasibility.

To evaluate engagement effectiveness in a tractable manner, we use a Friis-type
effective exposure model for the electromagnetic pulse incident on the target.
The resulting quantity should not be interpreted as the exact energy absorbed
by a specific internal electronic component of the UAV. Rather, it serves as an
effective received/intercepted pulse-energy metric at the target scale, while
unresolved victim-side coupling effects—including shielding, polarization
mismatch, airframe scattering, wiring/port coupling, and internal susceptibility—are absorbed into the empirical device-response parameters introduced later.

The received pulse energy is modeled using an effective-gain approximation, in
which the impact of pointing jitter is captured through the mean effective
transmit gain
\begin{equation}
\bar{G}=\mathbb{E}[G(\theta)]
=G_{0}\,\bigl(1+2k\sigma_{\theta}^{2}\bigr)^{-1/2},
\end{equation}
which is treated as a deterministic quantity for a given jitter variance
\(\sigma_\theta^2\). Let \(R(t)\) denote the slant range and
\(A_{\mathrm{atm}}(f,t)\) the atmospheric attenuation in dB. Then, the
instantaneous received-power proxy during one pulse is written as
\[
P_{r}(t)=P_{t}\,\bar{G}\,\left(\frac{\lambda}{4\pi R(t)}\right)^{2}
\,10^{-A_{\mathrm{atm}}(f,t)/10},
\qquad 0\le t\le T,
\]
and the corresponding effective pulse-energy proxy is
\begin{equation}
E_{r}(t)=P_{r}(t)\,\tau_{p}.
\label{eq: Er(t)}
\end{equation}

Taking logarithms yields
\begin{equation}
\ln E_{r}(t)
=
\ln\!\bigl(P_t\tau_p\bar G\bigr)
-2\ln\!\left(\frac{4\pi}{\lambda}\right)
-2\ln R(t)
-\frac{\ln 10}{10}A_{\mathrm{atm}}(f,t).
\label{eq:LnE_corrected}
\end{equation}

Under the moment-matched approximations developed earlier, we model
\[
\ln R(t)\sim \mathcal{N}\!\left(\mu_{\ln R},\sigma_{\ln R}^{2}\right),
\qquad
A_{\mathrm{atm}}(f,t)\sim \mathcal{N}\!\left(\mu_A,\sigma_A^2\right).
\]
Since \(\bar G\) is deterministic in this effective-gain formulation, the log-energy is approximated as Gaussian:
\begin{equation}
\ln E_r(t)\sim \mathcal{N}\!\left(\mu_{\ln E},\sigma_{\ln E}^2\right),
\label{eq:LnEr_corrected}
\end{equation}
with
\begin{gather}
\mu_{\ln E}
=
\ln\!\bigl(P_t\tau_p\bar G\bigr)
-2\ln\!\left(\frac{4\pi}{\lambda}\right)
-2\mu_{\ln R}
-\frac{\ln 10}{10}\mu_A,
\\
\sigma_{\ln E}^{2}
=
4\sigma_{\ln R}^{2}
+
\left(\frac{\ln 10}{10}\right)^2\sigma_A^2.
\label{eq:sigma_lnE_corrected}
\end{gather}

Therefore, in the present formulation, the effect of pointing jitter enters through the deterministic mean effective gain \(\bar G\), while the stochastic spread of the received pulse energy is induced by random slant range and atmospheric attenuation.

The quantity \(E_r\) in \eqref{eq: Er(t)} is therefore interpreted as an effective exposure metric rather than the exact absorbed energy of a particular electronic subcomponent. The mapping from \(E_r\) to model-predicted engagement-effectiveness probability captures the net effect of unresolved target-side coupling and vulnerability through the empirical susceptibility parameters \(E_{\mathrm{th}}\) and \(\beta_{\kappa}\).
Experimental studies in \cite{Benford2025} have shown that UAV platforms and their onboard electronics can suffer functional disruption or permanent damage when exposed to sufficiently strong high-power microwave pulses. For commercial off-the-shelf (COTS) UAVs, the effective disruption threshold is often on the order of \(E_{\mathrm{th}}\approx 10^{-2}\,\mathrm{J}\), whereas military-grade or hardened UAVs may exhibit significantly greater tolerance due to shielding, surge protection, and subsystem redundancy. In the present framework, \(E_{\mathrm{th}}\) is treated as a tunable effective susceptibility parameter that subsumes unresolved target-side coupling and vulnerability effects. It is varied to represent different classes of UAV threats, ranging from \(10^{-2}\,\mathrm{J}\) for highly vulnerable targets to
\(1\,\mathrm{J}\) for more strongly hardened systems.

\noindent\textit{Literature-informed surrogate calibration of the susceptibility curve:}
In the absence of standardized public pulse-by-pulse susceptibility datasets for a common UAV victim class, the parameters \(E_{\mathrm{th}}\) and \(\beta_{\kappa}\) are not claimed to be identified from a dedicated experimental calibration campaign. Instead, they are selected through a literature-informed surrogate calibration. Specifically, the midpoint of the susceptibility curve is anchored to the representative disruption-threshold range reported in the available HPM/UAV literature, while the slope parameter \(\beta_{\kappa}\) is chosen so that the \(10\%\)--\(90\%\) transition width on the log-energy axis is consistent with the spread implied by those reported threshold values. Accordingly, the resulting susceptibility curve should be interpreted as a phenomenological, literature-anchored response model rather than as a victim-specific, experimentally fitted calibration.

To capture the probabilistic nature of target susceptibility due to manufacturing variance, shielding quality, and angle-of-incidence effects, we model the exposure-based engagement-effectiveness probability using a sigmoid (logistic) function of the received pulse energy $E_r$:
 
\begin{equation}
P_{\mathrm{kill}}(E_{r})
=\frac{1}{1+\exp\!\left[-\beta_{\kappa}\left(\ln E_{r}-\ln E_{\mathrm{th}}\right)\right]},
\qquad \beta_{\kappa}>0,
\label{eq:Pk_lnE}
\end{equation}
where $\beta_{\kappa}$ controls the slope of the transition on the log-energy axis. Equivalently, 
\[
P_{\mathrm{kill}}(E_r)=\left(1+\left(E_{\mathrm{th}}/E_r\right)^{\beta_\kappa}\right)^{-1}.
\]

Although the notation $P_{\mathrm{kill}}$ is retained for compactness and consistency with common defence-effectiveness terminology, it should be interpreted here as a model-predicted exposure-response probability under the adopted susceptibility assumptions, rather than as a victim-specific, experimentally validated kill probability.

If a representative lower-to-upper disruption interval \((E_{10},E_{90})\) is available in the literature, then a convenient surrogate calibration is possible.
\[
E_{\mathrm{th}}=\sqrt{E_{10}E_{90}},
\qquad
\beta_\kappa=\frac{2\ln 9}{\ln(E_{90}/E_{10})},
\]
so that the logistic susceptibility model matches the implied \(10\%\)--\(90\%\) transition width on the log-energy axis.
This function reflects the fact that small deviations around the threshold $E_{\mathrm{th}}$ can produce widely different outcomes, depending on the drone's internal hardware resilience and system uncertainties.

The mean per-pulse engagement-effectiveness probability is obtained by averaging the logistic device-response over the log-energy distribution of a
single pulse. 
Let $x=\ln E_{r}$. Under the log-normal approximation in \eqref{eq:LnEr_corrected},
we have $x\sim\mathcal{N}\!\left(\mu_{\ln E},\sigma_{\ln E}^{2}\right)$.
The mean per-pulse engagement-effectiveness probability is therefore:
\begin{equation}
\bar{P}_{\mathrm{kill}}
=\frac{1}{\sigma_{\ln E}\sqrt{2\pi}}
\int_{-\infty}^{\infty}
\frac{\exp\!\left[-\frac{(x-\mu_{\ln E})^{2}}{2\sigma_{\ln E}^{2}}\right]}
{1+\exp\!\left[-\beta_{\kappa}\left(x-\ln E_{\mathrm{th}}\right)\right]}
\,dx.
\label{eq:Pkill_mean_lnE}
\end{equation}
The integral in \eqref{eq:Pkill_mean_lnE} can be evaluated efficiently using Gauss–Hermite (GH) quadrature.
Using the affine change of variables
\begin{equation}
x=\mu_{\ln E}+\sqrt{2}\,\sigma_{\ln E}\,z,
\end{equation}
and using the $n$-point GH nodes/weights \(\{(z_i,w_i)\}_{i=1}^n\), we obtain
\begin{equation}
\bar{P}_{\mathrm{kill}}
\approx\frac{1}{\sqrt{\pi}}
\sum_{i=1}^{n}w_{i}\;
\frac{1}{1+\exp\!\left[-\beta_{\kappa}\left(\mu_{\ln E}+\sqrt{2}\,\sigma_{\ln E}\,z_{i}-\ln E_{\mathrm{th}}\right)\right]}.
\label{eq:pkill_gh_lnE}
\end{equation}

In this work, we use $n=10$ GH nodes as the default accuracy–complexity trade-off.
To substantiate this choice, Sec.~\ref{subsec:GH_convergence} reports a convergence
study versus $n$ and $\sigma_{\ln E}$ by comparing GH predictions against both
a high-order GH reference and Monte Carlo (MC) estimates over
$\sigma_{\ln E}\in[0.1,0.6]$.


Let $\Delta \triangleq 1/f_{\mathrm{PRF}}$ denote the inter-pulse spacing. The
independence approximation is valid only when $\Delta$ exceeds the correlation
times of the dominant slow fluctuations (e.g., range and pointing-jitter induced gain),
so that successive pulses effectively experience independent channel states. 
When $\Delta$ is not large compared to these correlation times, successive pulses are
correlated, and the i.i.d. expression becomes optimistic. To quantify this effect, we use an effective number of independent pulses $N_{\mathrm{eff}}\le N$ and write
\begin{equation}
P_{\mathrm{kill,tot}}(T)\approx 1-\bigl(1-\bar{P}_{\mathrm{kill}}\bigr)^{\,N_{\mathrm{eff}}},\qquad N=f_{\mathrm{PRF}}T.
\label{eq:Pkill_tot}
\end{equation}
A standard approximation for correlated samples expresses $N_{\mathrm{eff}}$ via the
normalized autocorrelation $\rho_k$ at lag $k$ as follows:
\begin{equation}
N_{\mathrm{eff}}\approx
\frac{N}{\,1+2\sum_{k=1}^{N-1}\left(1-\frac{k}{N}\right)\rho_k\,}.
\label{eq:Neff_general}
\end{equation}
For an AR(1) correlation model $\rho_k=\rho^{k}$ (with $0\le\rho<1$), this reduces, for
large $N$, to the closed-form approximation
\begin{equation}
N_{\mathrm{eff}}\approx N\,\frac{1-\rho}{1+\rho},
\label{eq:Neff_ar1}
\end{equation}
where $\rho$ may be related to a correlation time $T_c$ by $\rho \approx \exp(-\Delta/T_c)$.

In simulations, $\rho$ is estimated from the sample autocorrelation of $\ln E_r[n]$ or from range/jitter state trajectories and then mapped to $N_{\mathrm{eff}}$ via \eqref{eq:Neff_general}. The i.i.d. case is recovered for $\rho_k=0$.

The quantity \(P_{\mathrm{kill,tot}}(T)\) in \eqref{eq:Pkill_tot} is interpreted as a cumulative model-predicted engagement-effectiveness measure under the adopted propagation, pulse-correlation, and susceptibility assumptions. Here, $\bar{P}_{\mathrm{kill}}$ is the mean per-pulse exposure-response probability defined previously. Although the notation follows common defence-effectiveness terminology, this expression consolidates the effects of the transmitter and tracking parameters $\{P_{t},D,\sigma_{\theta},\tau_{p},f_{\mathrm{PRF}}\}$, the stochastic channel variables $\{R(t),A_{\mathrm{atm}}(f,t)\}$, and the susceptibility-response parameters $\{E_{\mathrm{th}},\beta_{\kappa}\}$ into a single probabilistic performance metric, thereby enabling parametric assessment, trade-off analysis, and sensitivity studies for HPM counter-UAV system design.

\section{\label{sec:Sensitivity-Analysis-of}Sensitivity Analysis of Received Energy and System Parameters}

This section quantifies how variations in key system and environmental parameters influence the received pulse energy and, consequently, the model-predicted engagement-effectiveness probability. A sensitivity framework is developed to identify the most influential parameters, providing a principled basis for HPM counter-UAV system sizing, trade-off analysis, and risk-aware mission planning.

\subsection{\label{subsec: 2.1}Received-energy model for sensitivity analysis}

As established in Eq. \eqref{eq: Er(t)}, the received energy per pulse, $E_{r}(t)$, is a function of the transmit power, pulse width, antenna gain (including pointing jitter), stochastic range, and atmospheric attenuation. For sensitivity analysis, we focus on the statistical mean, $\bar{E}$,
which can be expressed as
\begin{gather}
\bar{E}=\mathbb{E}\left[E_{r}(t)\right]=P_{t}\,\tau_{p}\,G_{0}\,(1+2k\,\sigma_{\theta}^{2})^{-\frac{1}{2}}\,\left(\frac{\lambda}{4\pi}\right)^{2}\nonumber \\
\exp\left(-2\mu_{\ln R}+2\sigma_{\ln R}^{2}\right)\,\exp\left(-\frac{\ln10}{10}\,\mu_{A}+\frac{1}{2}\left(\frac{\ln10}{10}\right)^{2}\sigma_{A}^{2}\right).\label{eq: E(Er)}
\end{gather}
where $\sigma_{\theta}^{2}$ is the pointing jitter variance, $\left(\mu_{\ln R},  
 \sigma_{\ln R}^{2}\right)$ represents the mean and variance of the log-range, respectively, and $\mu_{A}$  is the mean atmospheric attenuation in dB. Equation \eqref{eq: E(Er)}
provides the exact expectation whenever attenuation is treated as a random (log-normal) variable; if $\sigma_{A}^{2}\approx0$ represents negligible variability, the final exponential reduces to the familiar deterministic factor
$10^{-\mu_{A}/10}.$

\subsection{Normalized Sensitivity Definition}

To assess how variations in each parameter propagate to the expected received energy, we introduce the normalized sensitivity coefficient (also known as elasticity), defined as:
\begin{equation}
S_{\chi}=\frac{\partial\ln\bar{E}}{\partial\ln\chi}.\label{eq: S_X}
\end{equation}
Here, $S_{\chi}$ quantifies the percentage change in the mean received energy $\bar{E}$ resulting from a one percent change in the parameter $\chi$, where $\chi$ denotes any relevant system or channel variables, such as transmit power, pulse width, antenna diameter, pointing jitter, mean range, or atmospheric attenuation.

By differentiating the analytical expression for $\bar{E}$ in \eqref{eq: E(Er)} with respect to the natural logarithm of each parameter, we obtain the explicit closed-form sensitivities summarized in Table~\ref{tab:sensitivities}.
For the aperture diameter $D$ and the pointing-jitter standard deviation $\sigma_{\theta}$, 
the normalized sensitivities (defined as in~(\ref{eq: S_X})) are
\begin{gather}
S_{D}=2-\frac{2k\sigma_{\theta}^{2}}{1+2k\sigma_{\theta}^{2}},\label{eq:S-D}\\
S_{\sigma_{\theta}}=-\frac{2k\sigma_{\theta}^{2}}{1+2k\sigma_{\theta}^{2}},\label{eq: S_f}
\end{gather}
where all terms retain their definitions from Section \ref{subsec: 2.1}.
These expressions establish an explicit quantitative relationship between system design parameters, such as aperture diameter $D$ and transmit power $P_{t}$, and stochastic factors, including pointing jitter and atmospheric variability, thereby enabling rigorous prediction of  HPM  counter-UAV engagement performance.
\par\vspace{0.4\baselineskip}

\begin{table}[t]
\centering
\caption{Normalized sensitivities of the mean received pulse energy (log-domain).}
\label{tab:sensitivities}
\begin{tabular}{l c}
\hline
\textbf{Sensitivity} & \textbf{Value} \\
\hline
$S_{P_{t}}$            & $1$ \\
$S_{\tau_{p}}$         & $1$ \\
$S_{\eta}$             & $1$ \\
$S_{\bar{R}}$          & $-2$ \\
$S_{\mu_{A}}$          & $-\tfrac{\ln 10}{10}\approx -0.23$ \\
$S_{\sigma_{A}^{2}}$   & $\tfrac{1}{2}\!\left(\tfrac{\ln 10}{10}\right)^{2}\approx 0.0265$ \\
\hline
\end{tabular}
\end{table}

\par\vspace{0.4\baselineskip}
In particular, the results indicate that the received energy exhibits a consistently linear dependence on, and high sensitivity to, variations in transmit power and pulse duration. The sensitivity to antenna diameter is strongly modulated by pointing jitter. In the limit of negligible jitter, increasing the antenna diameter yields a quadratic enhancement in received energy; however, in the jitter-dominated regime this benefit is effectively suppressed. Pointing jitter, mean range, and mean atmospheric attenuation all contribute negative sensitivities, with range producing the most pronounced decline owing to its inverse-square scaling. The influence of fluctuations in atmospheric attenuation is comparatively small; nevertheless, it is included in the analysis for completeness.

\section{\label{sec:Results }Results and Discussion}

To verify the analytical framework developed in Section~\ref{sec:Sys-Mod}, we compare the closed-form predictions with Monte Carlo (MC) simulations generated under the same modelling assumptions. The purpose of this comparison is to confirm the internal consistency and numerical accuracy of the proposed probabilistic formulation, rather than to claim victim-specific experimental validation. Unless otherwise specified, the reference scenario uses the parameters listed in Table~\ref{tab:simparams}. These values are selected to be representative of contemporary HPM systems and are informed by publicly available information on platforms such as AFRL's THOR, CHAMP, and Raytheon's PHASER \cite{Benford2025,DefenseUpdate2019,Laboratory2023}, which employ high peak powers and microsecond-scale pulses.

The baseline atmospheric results should be interpreted under the finite-layer slant-path approximation introduced in Sec.~II-C. In this formulation, rain attenuation is governed by the in-rain path length $L_{\mathrm{rain}}(t)$ and the corresponding effective attenuation moments $(\mu_A,\sigma_A^2)$, which are then adjusted to emulate different weather severities and propagation conditions.

Uncertainty in the model-predicted effectiveness measures, \(\bar P_{\mathrm{kill}}\) and \(P_{\mathrm{kill,tot}}\), may be assessed through bootstrap resampling of Monte Carlo simulations and parametric perturbation of \((E_{\mathrm{th}},\beta_\kappa)\).

\begin{table}[h]
\centering
\caption{Baseline simulation parameters for HPM--UAV engagement}
\label{tab:simparams}
\setlength{\tabcolsep}{4pt}
\renewcommand{\arraystretch}{0.92}
\begin{tabular}{lcc}
\hline
\textbf{Parameter} & \textbf{Symbol} & \textbf{Baseline Value} \\
\hline
Peak transmit power & $P_{t}$ & $200$ kW \\
Pulse width & $\tau_{p}$ & $0.5~\mu$s \\
Pulse repetition frequency & $f_{\mathrm{PRF}}$ & $1$ kHz \\
Carrier frequency & $f$ & $2.45$ GHz (ISM band) \\
Wavelength & $\lambda$ & $0.122$ m \\
Antenna diameter & $D$ & $1.5$ m \\
Pointing jitter (std. dev.) & $\sigma_{\theta}$ & $1$ mrad \\
Slant range (nominal) & $\bar{R}$ & $1$ km \\
Mean specific attenuation (baseline) & $\bar{\gamma}_{\mathrm{atm}}$ & $0.2$ dB/km \\
Integrated attenuation (baseline, at $\bar{R}=1$ km) & $\mu_{A}$ & $0.2$ dB \\
Effective susceptibility threshold & $E_{\mathrm{th}}$ & $10^{-2}$ J \\
Susceptibility-response slope & $\beta_{\kappa}$ & $50$ \\
\hline
\end{tabular}
\end{table}

\subsection{\label{subsec:GH_convergence}Convergence of Gauss--Hermite Quadrature}

The mean per-pulse engagement-effectiveness probability \(\bar{P}_{\mathrm{kill}}\) is evaluated using the
\(n\)-point Gauss–Hermite (GH) quadrature in \eqref{eq:pkill_gh_lnE}, based on the
log-normal received-energy model in \eqref{eq:LnEr_corrected} and the susceptibility model in \eqref{eq:Pk_lnE}. Here, \(n\) denotes the number of GH nodes.

Although GH quadrature converges rapidly for smooth integrands, its accuracy depends on the log-energy spread \(\sigma_{\ln E}\), the susceptibility slope \(\beta_\kappa\), and the offset between the mean log-energy and the threshold. We therefore define
\begin{equation}
\delta \triangleq \mu_{\ln E}-\ln E_{\mathrm{th}},
\label{eq:delta_def}
\end{equation}
so that the numerical difficulty depends on \((\beta_\kappa,\sigma_{\ln E},\delta)\).

\textit{Verification protocol:} We assess GH accuracy by comparing \(\bar{P}_{\mathrm{kill}}^{(n)}\)
against: (i) a high-order GH reference value \(\bar{P}_{\mathrm{kill}}^{(\mathrm{ref})}\), computed using \(n_{\mathrm{ref}}\) nodes, and
(ii) a Monte Carlo (MC) estimate \(\bar{P}_{\mathrm{kill}}^{(\mathrm{MC})}\), obtained by sampling
\(x=\ln E_r \sim \mathcal{N}(\mu_{\ln E},\sigma_{\ln E}^2)\) and averaging \(P_{\mathrm{kill}}(e^x)\).
For each \((n,\sigma_{\ln E})\), we report

\begin{equation}
\epsilon_{\mathrm{GH}}(n,\sigma_{\ln E})=\left|\bar{P}_{\mathrm{kill}}^{(n)}-\bar{P}_{\mathrm{kill}}^{(\mathrm{ref})}\right|,\;\epsilon_{\mathrm{MC}}(n,\sigma_{\ln E})=\left|\bar{P}_{\mathrm{kill}}^{(n)}-\bar{P}_{\mathrm{kill}}^{(\mathrm{MC})}\right|.\label{eq:eps_defs}
\end{equation}

Table~\ref{tab:GHconv} reports these errors for representative spreads
\(\sigma_{\ln E}\in\{0.1,0.3,0.6\}\) and node counts \(n\in\{6,8,10,12\}\), using a representative offset \(\delta\) and a moderate susceptibility slope \(\beta_\kappa\).
The results show the dependence of GH convergence on both \(n\) and \(\sigma_{\ln E}\), supporting the use of \(n=10\) as a practical default for the representative regime considered here.

\begin{table}[t]
\centering
\caption{Convergence of \(n\)-point GH quadrature in \eqref{eq:pkill_gh_lnE} for evaluating \(\bar{P}_{\mathrm{kill}}\). Here \(n\) is the GH node count, \(n_{\mathrm{ref}}=80\) is the high-order GH reference node count, and \(N_s=10^6\) is the number of MC samples. The table uses \(\delta=-0.4\) and \(\beta_\kappa=6\).}
\label{tab:GHconv}
\renewcommand{\arraystretch}{1.15}
\begin{tabular}{c c c c}
\hline
\(\sigma_{\ln E}\) & \(n\) & \(\epsilon_{\mathrm{GH}}\) & \(\epsilon_{\mathrm{MC}}\) \\
\hline
0.1 & 6  & 3.684e-08 & 2.392e-05 \\
0.1 & 8  & 2.175e-10 & 2.389e-05 \\
0.1 & 10 & 2.232e-11 & 2.389e-05 \\
0.1 & 12 & 9.727e-13 & 2.389e-05 \\
\hline
0.3 & 6  & 7.947e-04 & 6.916e-04 \\
0.3 & 8  & 1.152e-04 & 1.206e-05 \\
0.3 & 10 & 1.995e-05 & 1.231e-04 \\
0.3 & 12 & 2.170e-05 & 1.249e-04 \\
\hline
0.6 & 6  & 5.330e-03 & 4.990e-03 \\
0.6 & 8  & 2.090e-03 & 2.430e-03 \\
0.6 & 10 & 3.419e-03 & 3.759e-03 \\
0.6 & 12 & 2.938e-03 & 3.279e-03 \\
\hline
\end{tabular}
\end{table}

\begin{figure}[t]
\centering
\includegraphics[width=0.88\linewidth]{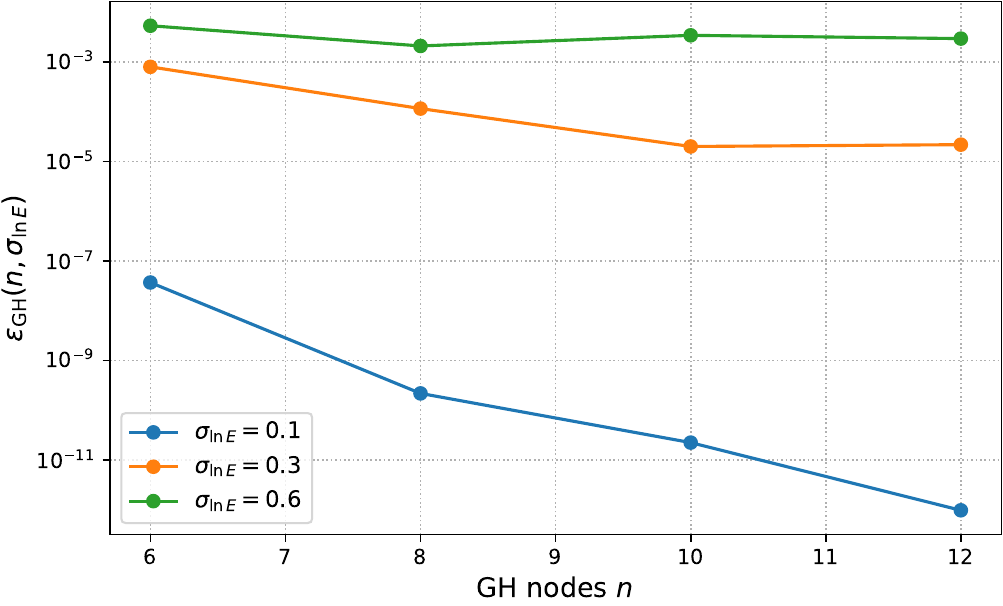}
\caption{GH quadrature convergence for \(\bar{P}_{\mathrm{kill}}\): absolute GH error \(\epsilon_{\mathrm{GH}}(n,\sigma_{\ln E})\) versus node count \(n\) for representative spreads \(\sigma_{\ln E}\in\{0.1,0.3,0.6\}\).}
\label{fig:Fig_ex}
\end{figure}

Fig.~\ref{fig:Fig_ex} confirms the numerical convergence of the GH approximation in \eqref{eq:pkill_gh_lnE}. As expected, the error decreases as \(n\) increases, while larger \(\sigma_{\ln E}\) leads to a more demanding integration regime. Together, Table~\ref{tab:GHconv} and Fig.~\ref{fig:Fig_ex} show that \(n=10\) is a reasonable default for the representative setting considered here, although steeper susceptibility transitions may require a larger node count.

\subsection{Propagation-level comparison with measurement-based references}
\label{subsec:ext_compare}

To further assess realism beyond direct analytical-versus-Monte-Carlo agreement, Table~\ref{tab:ext_compare} compares the proposed framework with representative measurement-based UAV air-to-ground references. Since openly available standardized end-to-end HPM engagement-effectiveness benchmarks for UAV targets remain very limited, the external comparison is performed mainly at the propagation-model level.

\begin{table*}[t]
\centering
\caption{Comparison of the proposed framework with representative measurement-based UAV air-to-ground references.}
\label{tab:ext_compare}
\renewcommand{\arraystretch}{1.15}
\begin{tabular}{>{\raggedright\arraybackslash}p{2.8cm}
                >{\raggedright\arraybackslash}p{4.0cm}
                >{\raggedright\arraybackslash}p{3.8cm}
                >{\raggedright\arraybackslash}p{3.8cm}}
\hline
\textbf{Aspect} &
\textbf{Our study} &
\textbf{Ref.~\cite{Lyu_TAP_2024_27GHz_Rural}} &
\textbf{Ref.~\cite{Ni_TITS_2024_U2G_Shadowing}} \\
\hline
Primary scope
& End-to-end probabilistic HPM--UAV engagement-effectiveness framework
& UAV air-to-ground channel measurement/modeling at 2.7~GHz
& UAV-to-ground path-loss and shadowing study with built-up-area and airframe effects \\

Propagation model
& FSPL + gaseous attenuation + rain attenuation + optional shadow fading
& Measurement-based A2G propagation reference
& Measurement-based path loss and shadow-fading reference \\

Band / scenario
& 2.45~GHz baseline HPM engagement
& 2.7~GHz rural fixed-wing UAV channel
& UWB / wideband UAV-to-ground channel with built-up-area effects \\

Large-scale fading realism
& Optional log-normal shadow-fading extension \(X_{\mathrm{SF}}\sim\mathcal{N}(0,\sigma_{\mathrm{SF}}^{2})\)
& Frequency-near A2G realism reference
& Supports shadow-fading and airframe effects \\

End-to-end HPM metrics
& Reports \(\bar P_{\mathrm{kill}}\) and \(P_{\mathrm{kill,tot}}\)
& Not reported
& Not reported \\
\hline
\end{tabular}
\end{table*}

Table~\ref{tab:ext_compare} shows that the selected external references are most useful as propagation-realism benchmarks rather than as complete end-to-end HPM engagement-effectiveness benchmarks. In particular, Ref.~\cite{Lyu_TAP_2024_27GHz_Rural} provides a frequency-near UAV air-to-ground measurement reference, while Ref.~\cite{Ni_TITS_2024_U2G_Shadowing} motivates the inclusion of shadow fading and airframe/environment-induced variability. Accordingly, these references support the realism of the propagation component of the proposed framework, while the end-to-end engagement-effectiveness metrics remain specific to the present analytical model.

\subsubsection{Measurement-informed shadow fading (SF)}
\label{subsec:realism_checks}
In many low-altitude air-to-ground environments, large-scale shadow fading due
to terrain, buildings, and airframe effects is well modeled as a log-normal
multiplicative loss. We therefore augment the link loss in dB by an additional
term
\[
X_{\mathrm{SF}}\sim \mathcal{N}(0,\sigma_{\mathrm{SF}}^{2}),
\]
assumed to be independent of the short-time scale jitter and kinematic fluctuations.
Under this standard extension, \(\ln E_r\) remains Gaussian, and the variance
parameter of the log-energy model is updated as
\begin{equation}
\sigma_{\ln E}^{2}\leftarrow
\sigma_{\ln E}^{2}
+\left(\frac{\ln 10}{10}\right)^{2}\sigma_{\mathrm{SF}}^{2}.
\end{equation}
Thus, representative \(\sigma_{\mathrm{SF}}\) values reported in measurement-based UAV-to-ground studies can be incorporated directly into the same analytical framework, providing a tractable way to assess how shadow fading broadens the received-energy distribution and alters the predicted engagement-effectiveness probabilities.

\subsubsection{Antenna-pattern cross-check (optional when pattern data are available)}
To upper-bound the approximation error in the near-boresight gain mapping, we validate the Gaussian-based mean gain \(\mathbb{E}[G(\theta)]\) by comparing it with either (i) an Airy-pattern-based mean gain corresponding to a uniformly illuminated circular aperture or (ii) an empirically measured antenna pattern, in accordance with \eqref{eq:epsilonG_def}. This procedure provides a
systematic means to characterize the validity domain as a function of
\(\sigma_\theta\) and to ascertain whether the Gaussian main-lobe approximation remains sufficiently accurate for the pointing-jitter levels of interest.

\subsection{Effect of Susceptibility Threshold on Per-Pulse Engagement Effectiveness}

\begin{figure}
\begin{centering}
\includegraphics[width=3.5in,viewport=2bp 0bp 550bp 350bp]{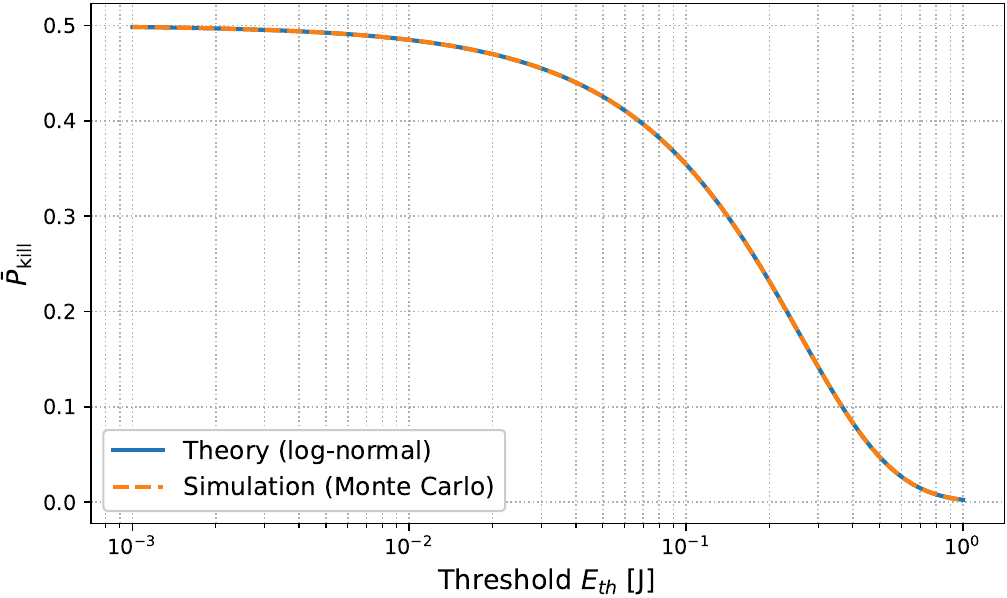}
\par\end{centering}
\caption{Mean per-pulse engagement-effectiveness probability \texorpdfstring{$\bar{P}_{\mathrm{kill}}$}{P\_kill} versus effective susceptibility threshold \texorpdfstring{$E_{\mathrm{th}}$}{E\_th}. Theoretical predictions are compared with Monte Carlo simulations, showing close agreement across thresholds under the adopted assumptions.\label{fig:2}}
\end{figure}

Fig.~\ref{fig:2} shows how the mean per-pulse engagement-effectiveness probability, \(\bar{P}_{\mathrm{kill}}\), varies with the effective susceptibility threshold \(E_{\mathrm{th}}\). The solid curve represents the analytical prediction based on the log-normal energy distribution, while the dashed curve shows MC results. The analytical predictions closely match the Monte Carlo simulations across all thresholds, confirming the internal consistency of the log-normal approximation under the adopted assumptions. At low thresholds (\(E_{\mathrm{th}}\sim10^{-3}\)--\(10^{-2}\,\mathrm{J}\)), representative of lightly shielded commercial UAVs, the engagement-effectiveness probability remains high and relatively stable. As \(E_{\mathrm{th}}\) increases, \(\bar{P}_{\mathrm{kill}}\) falls steeply, reflecting the diminishing likelihood of a single pulse producing a strong exposure response. For hardened platforms (\(E_{\mathrm{th}}\geq10^{-1}\,\mathrm{J}\)), the per-pulse engagement-effectiveness probability approaches zero, indicating that individual pulses are unlikely to be effective against resilient electronics under the adopted assumptions.

\subsection{Cumulative Engagement Effectiveness versus Dwell Time}

\begin{figure}
\begin{centering}
\includegraphics[width=3.5in,viewport=2bp 0bp 550bp 350bp]{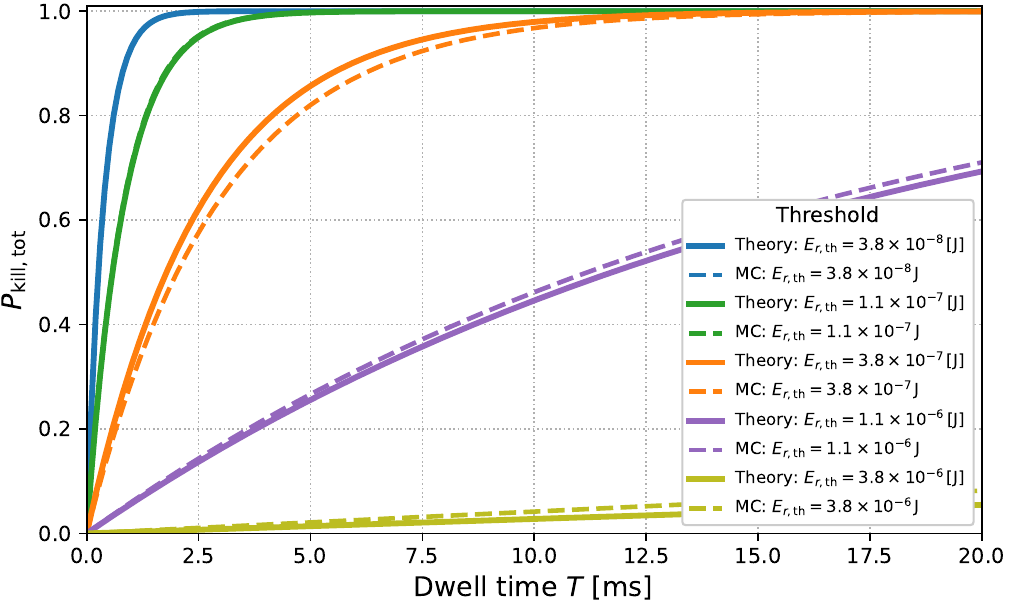}
\par\end{centering}
\caption{Cumulative engagement-effectiveness probability \(P_{\mathrm{kill,tot}}\) versus dwell time \(T\) for several effective susceptibility thresholds \(E_{\mathrm{th}}\). Theoretical predictions under the log-normal energy model and i.i.d. pulse assumption agree closely with Monte Carlo results. Higher thresholds slow the accumulation, while vulnerable targets approach unity within milliseconds.
\label{fig:3}}
\end{figure}

Fig.~\ref{fig:3} shows the cumulative engagement-effectiveness probability, \(P_{\mathrm{kill,tot}}\), as a function of dwell time for several effective susceptibility thresholds. For each threshold, both theoretical predictions and MC results are plotted. For vulnerable UAVs, cumulative engagement effectiveness quickly saturates near unity within milliseconds, indicating the feasibility of rapid engagement under the adopted assumptions. At higher thresholds, the cumulative probability rises more gradually and may remain below \(50\%\) even after extended dwell times, illustrating the difficulty of achieving high predicted effectiveness against hardened or shielded targets. The close agreement between the analytical and Monte Carlo results supports the internal consistency of the proposed log-normal energy model and the GH-based evaluation under the adopted pulse-statistics assumptions. It illustrates the sensitivity of engagement performance to dwell time and target susceptibility.

\subsection{Statistical Distribution of Received Energy}

\begin{figure}
\begin{centering}
\includegraphics[width=3.5in,viewport=2bp 0bp 550bp 350bp]{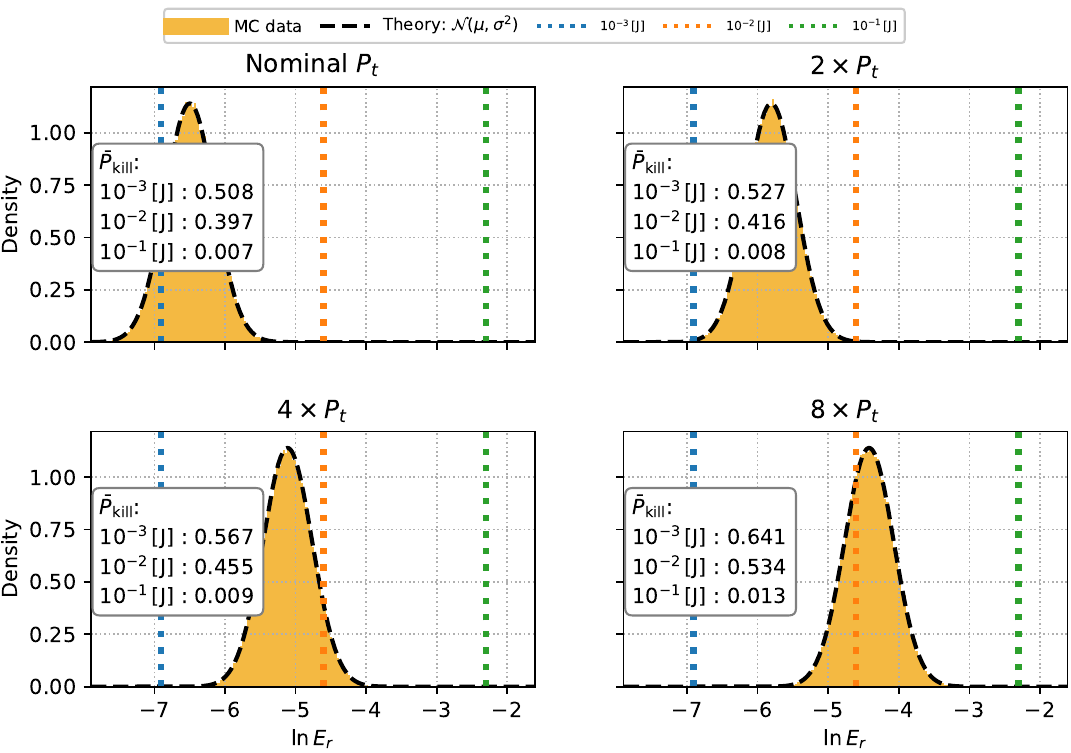}
\par\end{centering}
\caption{Effect of increasing $P_{t}$ on the distribution of received pulse
energy and mean engagement-effectiveness probability for log-normal parameters $\mu_{\ln E_r}=-6.5$ ,$\sigma_{\ln E_r}=0.35$ and logistic susceptibility-response slope $\beta_{\kappa}=50$ , evaluated for different $E_{\mathrm{th}}$ values.
\label{fig:4}}
\end{figure}

Fig.~\ref{fig:4} illustrates the distribution of received pulse energy \(\ln E_{r}\) under increasing transmit power levels \(P_t\) (from \(P_t\) to \(8P_t\)). Each subplot overlays an MC-derived histogram with the corresponding log-normal probability density function and marks three representative effective susceptibility thresholds (\(10^{-3}\,\mathrm{J}\), \(10^{-2}\,\mathrm{J}\), and \(10^{-1}\,\mathrm{J}\)) on the abscissa. Annotated boxes report the calculated values of \(\bar{P}_{\mathrm{kill}}\) for each threshold using Eq.~\eqref{eq:Pkill_mean_lnE}. The log-normal model accurately reproduces the MC-derived histograms. Increasing \(P_t\) shifts the energy distribution rightward while maintaining its log-normal shape, thereby increasing the fraction of pulses associated with higher exposure-response probability. For example, at the baseline power level, \(\bar{P}_{\mathrm{kill}}\) is approximately 0.508, 0.397, and 0.007 for the three thresholds; doubling \(P_t\) increases these values to 0.641, 0.534, and 0.013, respectively. This monotonic trend explains the dwell time results shown in Fig.~\ref{fig:3}: higher transmit power increases the per-pulse engagement-effectiveness probability, thus reducing the number of pulses, or dwell time, required to achieve a desired cumulative engagement-effectiveness probability.

\subsection{Per-Pulse Engagement Effectiveness versus Transmit Power}

\begin{figure}[t]
\begin{centering}
\includegraphics[width=3.3in,viewport=2bp 0bp 550bp 350bp]{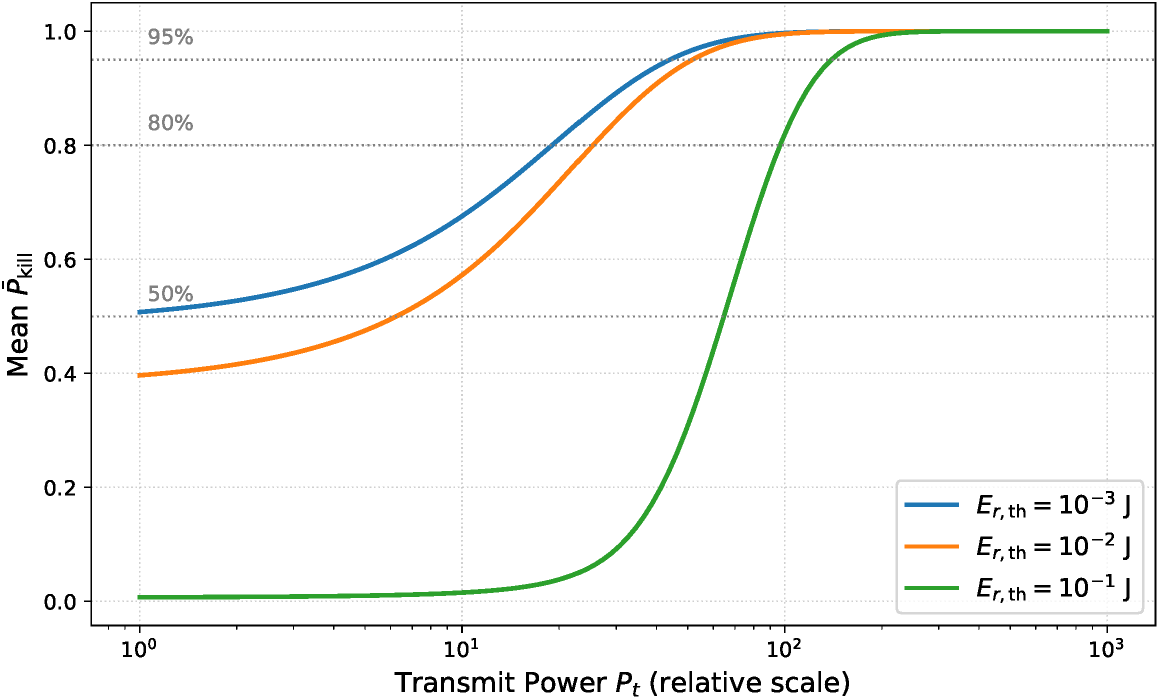}
\par\end{centering}
\caption{Mean per–pulse engagement-effectiveness probability $\bar{P}_{\mathrm{kill}}$ versus transmit power $P_t$ for multiple effective susceptibility thresholds $E_{\mathrm{th}}$ (theory: solid; MC: markers). Baseline unless noted: $\tau_p=0.5~\mu\mathrm{s}$, $f_{\mathrm{PRF}}=1~\mathrm{kHz}$, $f=2.45~\mathrm{GHz}$ ($\lambda=0.122~\mathrm{m}$), $D=1.5~\mathrm{m}$, $R=1~\mathrm{km}$, $\sigma_\theta=1~\mathrm{mrad}$, $\mu_A=0.2~\mathrm{dB}$, $\beta_\kappa=50$.}
\label{fig:5}
\end{figure}

Fig.~\ref{fig:5} shows the mean per-pulse engagement-effectiveness probability as a function of relative transmit power for three effective susceptibility thresholds. The abscissa is normalized to a reference transmit power and displayed logarithmically; while the ordinate shows $\bar{P}_{\mathrm{kill}}$. All curves exhibit a sigmoidal form, consistent with the logistic susceptibility-response model in Eq.~\eqref{eq:Pkill_mean_lnE}. For low thresholds, modest increases in $P_{t}$ rapidly drive $\bar{P}_{\mathrm{kill}}$ towards unity. In contrast, for hardened targets ($\mathrm{10^{-1}}$ J), substantial increases in power are required to achieve even a moderate engagement-effectiveness probability. These results indicate the strong influence of transmission power on the predicted engagement performance within the adopted model.

\noindent\textbf{Scope of the sensitivity ranking:}
The elasticities in Fig.~\ref{fig:8} and Table~\ref{tab:sensitivities} are evaluated at the baseline operating point
($f=2.45$~GHz, moderate $\mu_A$, and the engagement geometry considered here). At higher carrier frequencies and/or under severe rainfall,
$\mu_A$ and $\sigma_A^2$ increase, and the sensitivity contribution of atmospheric loss can become comparable to (or exceed) secondary factors
such as pointing jitter. Likewise, lower elevation angles increase the in-rain path length and can shift the ranking toward propagation-dominated behavior.
Therefore, the reported ranking should be interpreted as baseline-specific rather than universal.

\subsection{Role of Antenna Diameter}

\begin{figure}[t]
\begin{centering}
\includegraphics[width=3.5in,viewport=2bp 0bp 550bp 350bp]{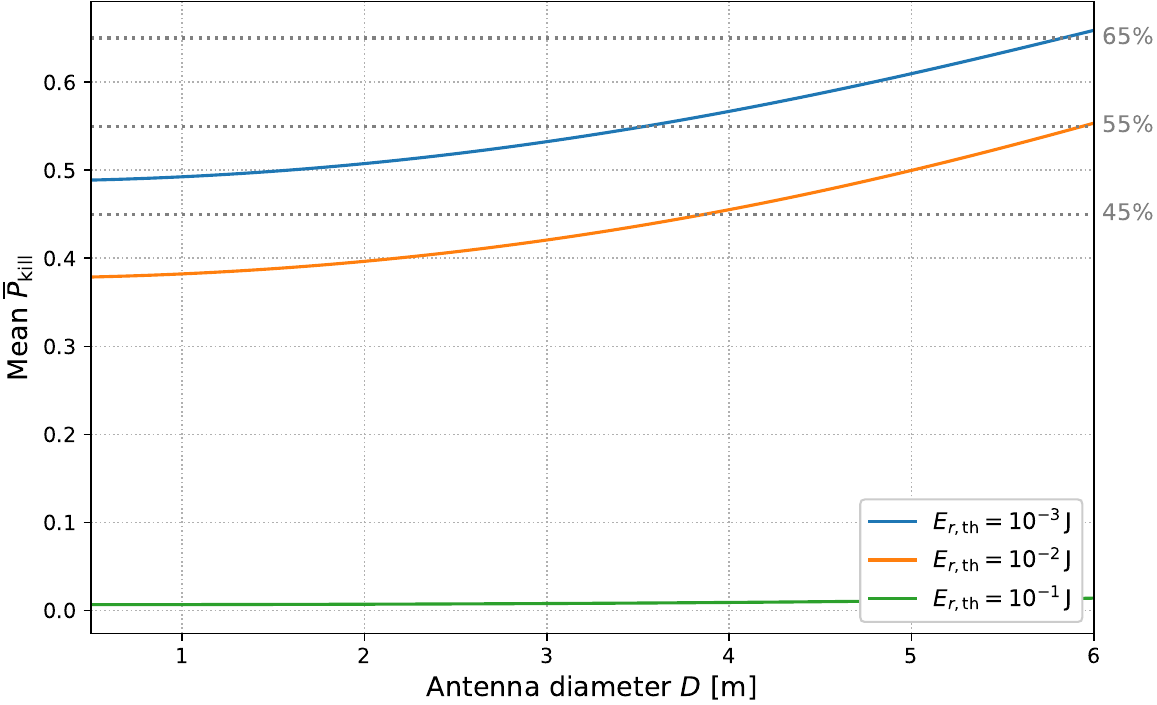}
\par\end{centering}
\caption{Mean per–pulse engagement-effectiveness probability $\bar{P}_{\mathrm{kill}}$ versus antenna diameter $D$ for multiple effective susceptibility thresholds $E_{\mathrm{th}}$ (theory: solid; MC: markers). The dependence on $D$ includes both the aperture gain $G_0=\eta(\pi D/\lambda)^2$ and $\theta_{3\mathrm{dB}}\approx 1.03\,\lambda/D$. Baseline unless noted: $P_t=200~\mathrm{kW}$, $\tau_p=0.5~\mu\mathrm{s}$, $f_{\mathrm{PRF}}=1~\mathrm{kHz}$, $f=2.45~\mathrm{GHz}$ ($\lambda=0.122~\mathrm{m}$), $R=1~\mathrm{km}$, $\sigma_\theta=1~\mathrm{mrad}$, $\mu_A=0.2~\mathrm{dB}$, $\beta_\kappa=50$.}
\label{fig:6}
\end{figure}

Fig.~\ref{fig:6} explores the dependence of \(\bar{P}_{\mathrm{kill}}\) on the antenna diameter \(D\) for three effective susceptibility thresholds. As expected, increasing \(D\) enhances antenna gain and, therefore, the received energy, leading to higher engagement-effectiveness probabilities. This improvement is most pronounced for low thresholds, where a moderate increase in aperture, e.g., from \(0.5\,\mathrm{m}\) to \(2\,\mathrm{m}\), can raise \(\bar{P}_{\mathrm{kill}}\) well above \(50\%\). For higher thresholds (\(\geq10^{-1}\,\mathrm{J}\)), however, the probability remains low across the evaluated diameter range, highlighting the limited benefit of aperture growth alone against hardened electronics. These results demonstrate that antenna scaling is most effective for vulnerable or commercial-grade UAVs, while its impact is more limited for hardened platforms.

\subsection{Dependence on Slant Range}

\begin{figure}[t]
\begin{centering}
\includegraphics[width=3.3in,viewport=2bp 0bp 550bp 350bp]{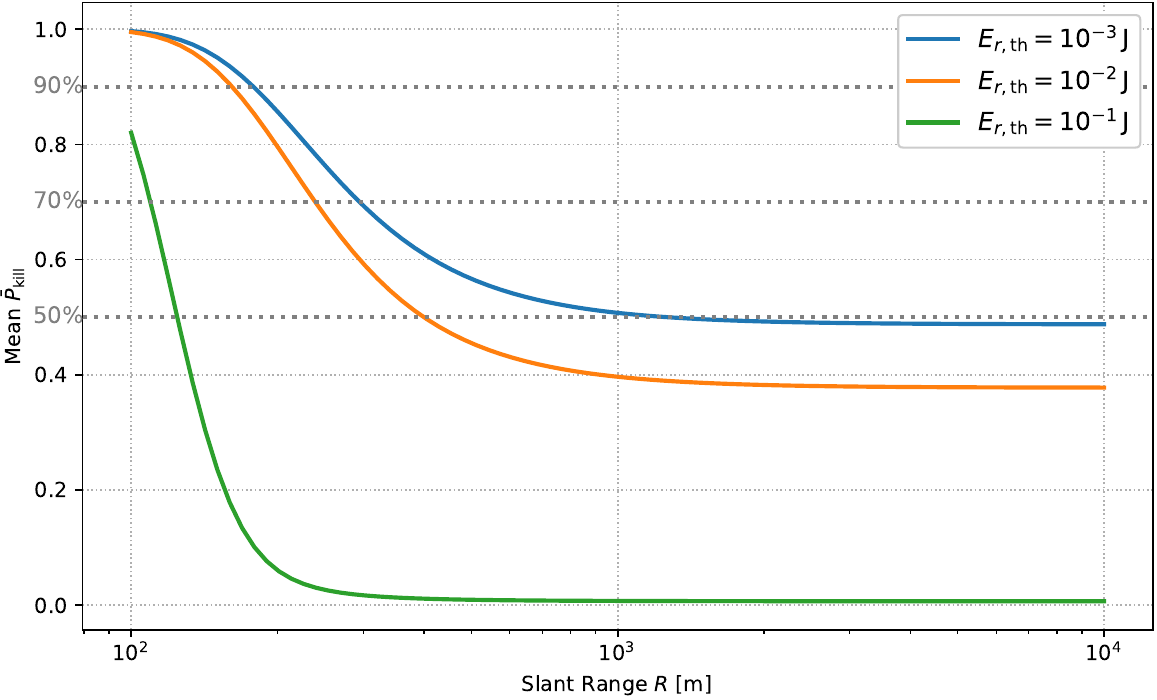}
\par\end{centering}
\caption{Mean per–pulse engagement-effectiveness probability $\bar{P}_{\mathrm{kill}}$ versus slant range $R$ for multiple effective susceptibility thresholds $E_{\mathrm{th}}$ (theory: solid; MC: markers). The dependence on $R$ reflects both free–space spreading $\propto(\lambda/4\pi R)^2$ and atmospheric attenuation $A_{\mathrm{atm}}(f,R)$, as modeled in Section~II. Baseline unless noted: $P_t=200~\mathrm{kW}$, $\tau_p=0.5~\mu\mathrm{s}$, $f_{\mathrm{PRF}}=1~\mathrm{kHz}$, $f=2.45~\mathrm{GHz}$ ($\lambda=0.122~\mathrm{m}$), $D=1.5~\mathrm{m}$, $\sigma_\theta=1~\mathrm{mrad}$, $\mu_A=0.2~\mathrm{dB/km}$, $\beta_\kappa=50$.}
\label{fig:7}
\end{figure}

The sensitivity of the per-pulse engagement-effectiveness probability \(\bar{P}_{\mathrm{kill}}\) to the engagement slant range \(R\) for three effective susceptibility thresholds is shown in Fig.~\ref{fig:7}. Across all thresholds, \(\bar{P}_{\mathrm{kill}}\) decreases approximately as \(R^{-2}\), which is consistent with free-space path loss. For the most vulnerable targets (\(E_{\mathrm{th}}=10^{-3}\,\mathrm{J}\)), the predicted engagement-effectiveness probability remains above \(50\%\) at approximately one kilometer under the baseline settings. By contrast, for the most resilient targets (\(E_{\mathrm{th}}=10^{-1}\,\mathrm{J}\)), the predicted engagement effectiveness becomes essentially negligible beyond a few hundred meters. This behavior highlights the fundamental trade-off between engagement range and target susceptibility, with direct implications for system positioning and operational planning.

\subsection{Sensitivity Analysis of Mean Received Energy}
\begin{figure}[t]
\begin{centering}
\includegraphics[width=3.1in,viewport=2bp 0bp 550bp 350bp]{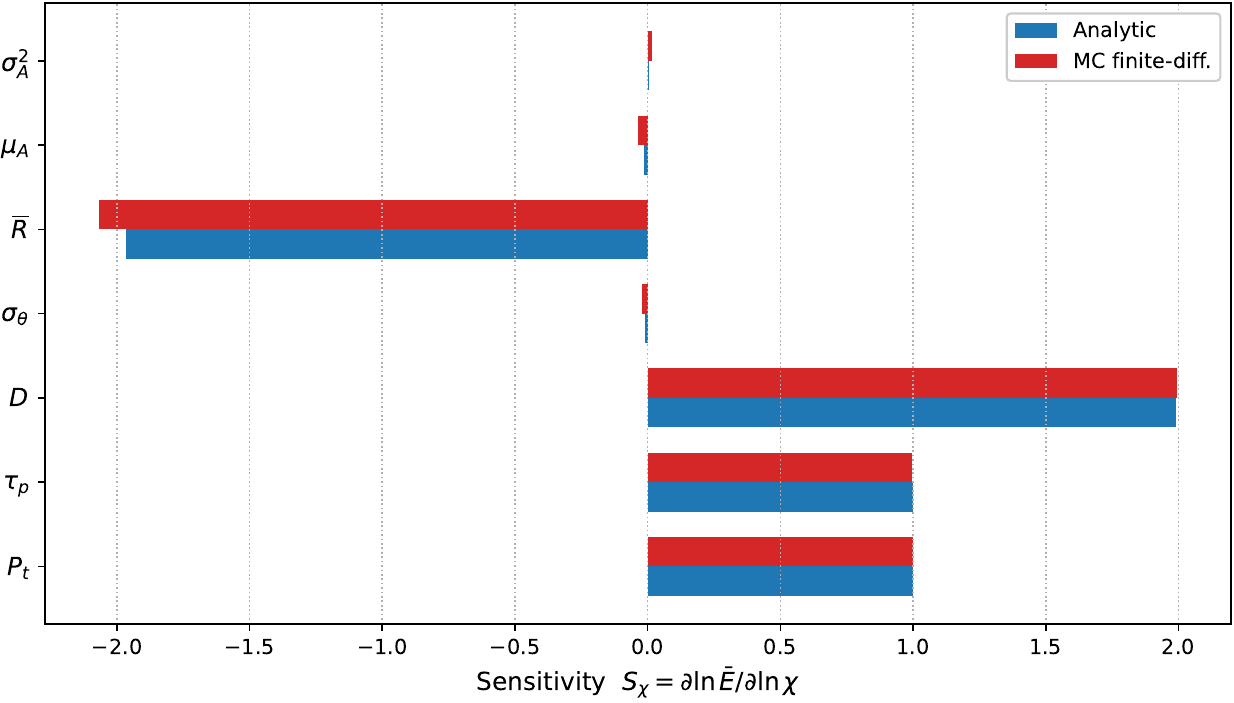}
\par\end{centering}
\caption{Normalized sensitivity (elasticity) of mean received energy $\bar{E}$ with respect to key system and channel parameters, $S_{\chi}\triangleq\partial\ln\bar{E}/\partial\ln\chi$. Bars/curves are evaluated from the analytic model of Section~III for $\chi\in\{P_t,\tau_p,D,\sigma_\theta,\bar{R},\mu_A,\sigma_A^2\}$ at the baseline operating point. Baseline unless noted: $P_t=200~\mathrm{kW}$, $\tau_p=0.5~\mu\mathrm{s}$, $f_{\mathrm{PRF}}=1~\mathrm{kHz}$, $f=2.45~\mathrm{GHz}$ ($\lambda=0.122~\mathrm{m}$), $D=1.5~\mathrm{m}$, $R=1~\mathrm{km}$, $\sigma_\theta=1~\mathrm{mrad}$, $\mu_A=0.2~\mathrm{dB/km}$. For reference, the closed‐form elasticities are $S_{P_t}=S_{\tau_p}=1$, $S_{\bar{R}}=-2$, $S_{\mu_A}=-\ln(10)/10$, and $S_{\sigma_A^2}=\tfrac{1}{2}\!\big(\ln(10)/10\big)^2$, while $S_{D}$ and $S_{\sigma_\theta}$ include the beam–jitter coupling through $k=4\ln 2/\theta_{3\mathrm{dB}}^{2}$ with $\theta_{3\mathrm{dB}}=1.03\lambda/D$.}
\label{fig:8}
\end{figure}

Fig.~\ref{fig:8} summarizes the normalized sensitivities \(S_{\chi}\)
(as defined in \eqref{eq: S_X}) of the mean received energy \(\bar{E}\) with
respect to the main system and channel parameters: transmit power \(P_t\),
pulse width \(\tau_p\), antenna diameter \(D\), pointing jitter
\(\sigma_\theta\), mean slant range \(\bar{R}\), mean atmospheric attenuation
\(\mu_A\), and attenuation variance \(\sigma_A^2\). Analytic sensitivities are
compared with finite-difference estimates obtained from Monte Carlo
simulations. Positive values indicate that increasing the parameter results
in an increase in \(\bar{E}\), while negative values indicate the opposite. The results show
that \(\bar{E}\) is most strongly sensitive to slant range, with
\(S_{\bar{R}}\approx -2\), indicating that a 1\% increase in range produces an
approximately 2\% decrease in mean energy. Transmit power, pulse width, and
antenna diameter exhibit positive sensitivities close to \(+1\) at the baseline
operating point, reflecting their strong direct influence on the link budget.
By contrast, pointing jitter, mean atmospheric attenuation, and attenuation variance have
smaller effects in the evaluated regime. The close agreement between the
analytical and Monte Carlo sensitivity estimates supports the internal
consistency of the theoretical sensitivity expressions and provides a useful
ranking of the dominant design levers at the baseline operating point. Within
this baseline setting, the largest gains are obtained by reducing range and by
increasing either aperture size or transmit power, whereas incremental
improvements in jitter or atmospheric compensation provide comparatively smaller
benefits.

\section{\label{sec:Conclusion}Conclusion}

This work presents a unified probabilistic framework for assessing the effectiveness of HPM engagements against UAVs under specified propagation, pointing, and susceptibility assumptions. The framework integrates (i) stochastic UAV kinematics with closed-form position and range statistics, (ii) an effective near-boresight beam-steering model that converts pointing jitter into a gain reduction law, (iii) an atmospheric channel model with tractable moments for gaseous and precipitation loss, and (iv) a device-response model that maps received pulse energy to the engagement-effectiveness probability via a logistic transition. A key contribution is the derivation of closed-form moments and log-normal closures that make the received-energy distribution and the associated per-pulse engagement-effectiveness probability analytically evaluable, substantially reducing reliance on Monte Carlo weather and geometry realizations while preserving statistical fidelity. A second contribution is an explicit sensitivity (elasticity) analysis that yields closed-form expressions for the dependence of the mean received energy on core design and environmental parameters, providing transparent levers for system optimization and mission planning. A third contribution is the end-to-end internal-consistency verification: analytical predictions closely agree with Monte Carlo simulations across a wide parameter range under the same modeling assumptions, supporting the numerical accuracy and practical utility of the framework.

The results show that the predicted engagement effectiveness is governed by the interplay between target susceptibility and the HPM link budget. The electronic vulnerability of the target, represented by the effective susceptibility threshold \(E_{\mathrm{th}}\), is the dominant factor, followed by transmit power, aperture size, its interaction with pointing jitter, and engagement range. Consistency between analysis and simulation supports the internal consistency of the received-energy and exposure-response probability models under the adopted assumptions. The closed-form sensitivities clarify why decreasing range or increasing either power or aperture yields the largest gains, whereas incremental improvements in pointing accuracy or atmospheric conditions provide comparatively modest benefits over the evaluated regimes. For commercial or unshielded UAVs, the model predicts that current HPM systems may achieve high cumulative engagement effectiveness within typical dwell times; for hardened platforms, the same framework indicates that comparable performance would require materially higher power, larger apertures, and/or longer dwell times, with the achievable range further constrained by propagation and pointing effects.

Overall, the presented framework constitutes a tractable and design-oriented analytical tool for probabilistic HPM--UAV engagement assessment within the scope of the adopted assumptions. It transforms stochastic kinematics, effective beam-control uncertainty, and atmospheric variability into closed-form performance metrics and sensitivity measures that enable rapid trade-off evaluation, parameter optimization, and risk-aware mission planning. Future research will extend the framework to incorporate time-correlated pulse trains, multi-target interactions, and adaptive scheduling strategies that exploit real-time estimates of range, pointing jitter, and atmospheric conditions.

\bibliographystyle{IEEEtran}
\bibliography{UAV}

\end{document}